\documentclass[a4paper,fleqn]{cas-dc}

\usepackage{amsmath}
\usepackage[numbers,sort&compress]{natbib}
\usepackage{xcolor}
\usepackage{hyperref}
\hypersetup{
colorlinks=true,
linkcolor=RoyalBlue,
citecolor=RoyalBlue,
urlcolor=RoyalBlue
}
\makeatletter
\pretocmd{\NAT@open}{\begingroup\color{\@citecolor}}{}{}
\apptocmd{\NAT@close}{\endgroup}{}{}
\makeatother
\def\tsc#1{\csdef{#1}{\textsc{\lowercase{#1}}\xspace}}
\tsc{WGM}
\tsc{QE}
\tsc{EP}
\tsc{PMS}
\tsc{BEC}
\tsc{DE}

\begin{document}
\let\WriteBookmarks\relax
\def\floatpagepagefraction{1}
\def\textpagefraction{.001}

\shorttitle{DeepLabv3++}


\title [mode = title]{Polyp segmentation in colonoscopy images using DeepLabV3++}                      



\author[1]{Al Mohimanul Islam}
\cormark[2]
\ead{E-mail address: aislam192054@gmail.com}

\author[1]{Sadia Shakiba Bhuiyan}
\cormark[2]
\ead{sbhuiyan192050@bscse.uiu.ac.bd}

\author[1]{Mysun Mashira}
\ead{mmashira201011@bscse.uiu.ac.bd}

\author[1,3]{Md. Rayhan Ahmed}
\ead{rayhan@cse.uiu.ac.bd}

\author[2]{Salekul Islam}
\cormark[1]
\ead{salekul.islam@northsouth.edu}

\author[1]{Swakkhar Shatabda}
\cormark[1]
\ead{sshatabda@gmail.com}

\affiliation[1]{organization={Department of Computer Science and Engineering, United International University},city={Dhaka},country={Bangladesh}}

\affiliation[2]{organization={Department of Electrical and Computer Engineering, North South University},city={Dhaka},country={Bangladesh}}

\affiliation[3]{organization={Department of Computer Science, University of British Columbia},city={Kelowna},country={Canada}}

\cortext[1]{Corresponding author.}
\cortext[2]{These two authors contributed equally to this work.}










\begin{abstract}
Segmenting polyps in colonoscopy images is essential for the early identification and diagnosis of colorectal cancer, a significant cause of worldwide cancer deaths. Prior deep learning based models such as Attention based variation, UNet variations and Transformer-derived networks have had notable success in capturing intricate features and complex polyp shapes. However they frequently encounter challenges in pinpointing small details and enhancing the representation of features on both local and global scale. In this study, we have introduced the DeepLabv3++ model which is an enhanced version of the DeepLabv3+ architecture. It is designed to improve the precision and robustness of polyp segmentation in colonoscopy images. We have utilized EfficientNetV2S within the encoder module for refined feature extraction with reduced trainable parameters. Additionally, we integrated Multi-Scale Pyramid Pooling (MSPP) and Parallel Attention Aggregation Block (PAAB) modules, along with a redesigned decoder, into our DeepLabv3++ model. The proposed model incorporates diverse separable convolutional layers and attention mechanisms within the MSPP block, enhancing its capacity to capture multi-scale and directional features. Additionally, the redesigned decoder further transforms the extracted features from the encoder into a more meaningful segmentation map. Our model was evaluated on three public datasets (CVC-ColonDB, CVC-ClinicDB, Kvasir-SEG) achieving Dice coefficient scores of 96.20\%, 96.54\%, and 96.08\%, respectively. The experimental analysis shows that DeepLabV3++ outperforms several state-of-the-art models in polyp segmentation tasks. Furthermore, compared to the baseline DeepLabV3+ model, our DeepLabV3++ with its MSPP module and redesigned decoder architecture,  significantly reduced segmentation errors (e.g., false positives/negatives) across small, medium, and large polyps. This improvement in polyp delineation is crucial for accurate clinical decision-making in colonoscopy.
\end{abstract}




\maketitle

\section{Introduction}

Colon cancer is a form of cancer that originates in the colon or rectum. It is also known as colorectal cancer \cite{paschke2018colon}. Colon cancer is one of the most deadly cancers and a major worldwide health concern \cite{LABIANCA2010106}.  It is the second most common cause of cancer-related deaths globally and the third most commonly diagnosed cancer. Considering occurrence rate, colon cancer ranks second in women and third in men \cite{bretthauer2022effect,xi2021global,sung2021global,amersi2005colorectal}. Colon and rectal cancer (CRC) starts as a little neoplastic polyp that grows over time and progresses through a stage known as dysplasia before becoming an invasive cancer. It is preventable if a colonoscopy identifies and removes it before it develops into a cancer cell \cite{ahmed2020colon,allen1995molecular,cappell2005colonic,sullivan2022cause}. A colonoscopy is crucial for identifying and preventing colon cancer and uses a flexible tube equipped with a camera to locate and remove polyps. However, the procedure takes a while and might overlook some polyps, so it needs to be carefully screened by experts manually \cite{gangrade2024modified}.  Manually identifying colonoscopy poses several challenges, like false polyp detection, as it highly depends on clinical experience and expertise. Undetected growth of polyps and false polyps can cause deadly colorectal cancer. They can also cause treatment delays that raise the risk of death \cite{giovannucci2002modifiable, ANANTHAKRISHNAN202016,ahmed2020colon}. Deep learning-based methods offer reliable, effective, and automated analysis of colonoscopy images, which has the potential to revolutionize the identification and treatment of colon cancer entirely. Automated segmentation techniques effectively analyze medical images with precision and objectivity, lessening the need for manual interpretation \cite{pacal2020comprehensive,tasnim2021deep,tharwat2022colon}.

Recent studies in deep learning for colonoscopy polyp segmentation have introduced numerous competitive models that strive to achieve high precision and real-time performance to enhance clinical usability. A recent research article proposed a modified DeeplabV3+ model which enhances DeeplabV3+ by incorporating attention mechanism, an Atrous Spatial Pyramid Pooling (ASPP) module, a ResNet101 backbone, and an encoder-decoder structure; the model successfully boosts the process of segmenting colonoscopy polyps \cite{gangrade2024modified}. Another study proposed two new approaches for colon polyp detection named TriUNet and DivergentNets. TriUNet is a segmentation model consisting of three distinct UNet models organized in a triangular configuration, and DivergentNets integrates the TriUNet framework with a collection of established segmentation models \cite{thambawita2021divergentnets}. The potential benefits of utilizing a deep learning-based CADx (Computer-Aided Detection and Diagnosis) system for colorectal cancer prevention include improving the efficiency and accuracy of identifying potential polyps during colonoscopy examinations. One of the studies proposed a CADx architecture TransResU-Net, which integrates transformer encoder blocks, residual blocks, a pre-trained ResNet50 model and dilated convolutions for real-time polyp segmentation \cite{tomar2022transresu}; another study explored DialatedSegNet which employs an architecture consisting of an encoder and decoder network. The encoder utilizes a pre-trained ResNet50 model to generate four distinct levels of feature maps. It can be incorporated into current CAD systems utilized for detecting colorectal cancer \cite{tomar2023dilatedsegnet}. HarDNet-MSEG, DRI-Net SegNet, and Deeplab neural networks are also some of the proposed models by recent studies for colonoscopy polyp segmentation using deep learning and CADx architectures \cite{huang2021hardnet,huang2018automatic,lan2023dri}.

The U-Net architecture is a convolutional neural network created for image segmentation tasks. It displays a symmetrical U-shaped framework consisting of an encoder for context extraction and a decoder for precise localization facilitated by skip connections \cite{ronneberger2015u}. It has gained significant prominence as a deep learning framework for segmenting medical images, with numerous variations and applications in various types of image formats \cite{punn2022modality,9446143}. The remarkable growth observed since 2017 showcases the significance and ability of medical image analysis to advance the field \cite{9446143}. One of the previous studies proposed an advanced neural network, A-DenseUNet, to enhance the precision of colorectal polyp identification in colonoscopy images \cite{safarov2021denseunet}. The model utilizes dilation convolution and an attention mechanism to capture wide-ranging perspectives without increasing computational expenses. In one of the previous studies, another modified U-net architecture called NeoUNet was proposed, where it performed well compared to other methods, especially in identifying cancerous polyps \cite{ngoc2021neounet}. NeoUNet's encoder-decoder model merges basic and advanced characteristics to separate objects accurately, improved by combining different loss functions. Enhanced U-Net models, which integrate atrous convolutions, attention mechanisms, or hybrid loss functions, improve polyp segmentation by capturing a broader context, eliminating extraneous data, and enhancing feature acquisition. These alterations allow for the exact outlining of boundaries and precise identification of high-risk cancerous growths, leading to a substantial enhancement in the quality of colonoscopy images \cite{ngoc2021neounet,raymann2022gar,safarov2021denseunet,thambawita2021divergentnets}.

In this paper we propose a novel method DeepLabV3++ which incorporates an EfficientNetV2S backbone for computationally efficient and precise polyp segmentation in colonoscopy images. We refine the existing DeepLabV3+ architecture by introducing a Multi-Scale Pyramid Pooling (MSPP) module and a Parallel Attention Aggregation Block (PAAB) to enhance the model's ability to capture complex polyp boundaries. Our contributions focus on these modifications and their impact on achieving this balance between efficiency and accuracy. To evaluate the effectiveness of our proposed approach, we conduct a comprehensive comparison with other leading polyp segmentation approaches. We present quantitative results including accuracy, precision, recall, and Intersection over Union (IoU) metrics. Additionally, we showcase qualitative examples demonstrating the model's improved performance in handling complex shapes and boundaries. Finally, we perform an ablation study to isolate the contributions of the MSPP and PAAB modules. The major contributions of this paper can be summarized as follows: 

\begin{itemize}
    \item We propose a robust segmentation framework for polyp segmentation tasks, based on the EfficientNetV2S encoder backbone and enhanced by an effective decoder module.
    \item We develop a novel multi-scale pyramid pooling block that utilizes diverse separable convolutions and kernel sizes. This method effectively extracts features from various scales of the feature maps, ensuring computational efficiency. Furthermore, the MSPP block includes skip connections to retain spatial information and improve gradient flow.
    \item We integrate a novel parallel attention aggregation block within the MSPP, designed to efficiently aggregate spatial and channel information from extracted features. The spatial attention mechanism in the PAAB utilizes multi-kernel separable convolutions, enhancing the model's spatial feature representation capabilities. Additionally, the channel attention mechanism enhances interdependencies among the extracted features.
    \item We conduct a comprehensive comparative analysis on three publicly available benchmark polyp datasets to evaluate the effectiveness of DeepLabV3++. Our method was compared against leading state-of-the-art polyp segmentation models, demonstrating it's significant performance improvements. 
\end{itemize}

The paper is organized as follows: In Section \ref{lit}, we reviewed related research. Our proposed DeepLabV3++ architecture is explained in Section \ref{method}. Next, in Section \ref{exp} we elaborated the experimental analysis of this study and Section \ref{discussion} presents key findings and results. Lastly, Section \ref{concl} concludes by summarizing our proposed model's architecture, improvements, outcomes, and future work.





\section{Related Work \label{lit}}
Many studies state that automated detection of polyps in colonoscopy improves the early identification of colorectal cancer through the provision of immediate alerts and enhances the accuracy of diagnosis \cite{xu2021real,lee2020real,luo2021artificial,nur2024automatic}. 

These studies explored various U-net variations and structures like hybrid methodology Convolution Neural Network (CNN) based methods, attention gates, and transformer-based methods for the segmentation of polyps in colonoscopy images, each utilizing distinct methodologies \cite{gangrade2024modified,tomar2022transresu,huang2022paformer,hosseinzadeh2020automatic,chen2021transunet}. In this section, we will review the methodologies utilized by these models.

\subsection{CNN based}

Qadir et al. have developed a real-time colon polyp detection system using single-shot feed-forward fully convolutional neural networks. The method achieved high accuracy on two polyp datasets, with 86.54\% recall, 86.12\% precision, and 86.33\% F1-score on the ETIS-LARIB dataset, and 91\% recall, 88.35\% precision, and 89.65\% F1-score on the CVC-ColonDB dataset \cite{qadir2021toward}.
In a study by Kassani et al., DenseNet169 outperformed other deep learning architectures for polyp segmentation in colorectal cancer screening with 99.15\% accuracy, 90.87\% Dice coefficient, and 83.82\% Jaccard index \cite{hosseinzadeh2020automatic}.
Nisha et al. developed a machine-learning algorithm to identify early-stage colorectal polyps from colonoscopy images. It outperformed other polyp detection methods \cite{nisha2022colorectal}.
In a study by Kassani et al., different deep learning architectures were compared for polyp segmentation in colorectal cancer screening. DenseNet169 outperformed other architectures with 99.15\% accuracy, 90.87\% Dice coefficient, and 83.82\% Jaccard index, showing significant improvement in polyp detection during colonoscopy \cite{hosseinzadeh2020automatic}.
Dash et al. developed an expert system using a Deep Belief Network to detect colorectal malignancy and identify polyp areas from complex images. The PLPNet system reduced mortality rates, improved prediction capabilities, and utilized bias-variance analysis and data augmentation to enhance accuracy \cite{dash2023identification}. 
HarDNet-MSEG is a convolutional neural network proposed by Huang et al. for polyp segmentation. It has achieved a 30\% inference time reduction compared to DenseNet and ResNet while achieving higher accuracy on ImageNet \cite{huang2101hardnet}.

\subsection{Transformer Based}

Zhang et al. introduced TransFuse, a method for medical image segmentation that combines CNNs and Transformers. It achieved state-of-the-art results on 2D and 3D medical image sets \cite{zhang2021transfuse}.
Shen et al. introduced the Convolution in Transformer (COTR) network for polyp detection. It's an end-to-end CAD system combining CNN for feature extraction, transformer encoder, and convolutional layers \cite{shen2021cotr}.
Wang et al. proposed the SSFormer, a state-of-the-art medical image segmentation model using a pyramid Transformer encoder for improved generalization. The model's robust generalization and accurate prediction abilities were demonstrated on the Kvasir and CVC-ClinicDB benchmarks \cite{wang2022stepwise}.
Nguyen et al. introduced the LAPFormer model, combining a hierarchical Transformer encoder and a CNN decoder to enhance polyp detectability and segmentation. It achieved competitive performance on all datasets while being less computationally expensive than other methods \cite{nguyen2022lapformer}.
Khaled et al. introduced the MA-NET model, which combined a modified Mix-ViT transformer and CLAHE preprocessing to improve colorectal polyp segmentation. The model excels at detecting small and flat polyps and was trained on Kvasir-SEG and CVC-ClinicDB datasets, with testing on ETIS-LaribPolypDB and CVC-ColonDB datasets, emphasizing the importance of high-quality training data for best results \cite{elkarazle2023improved}.
Zhiyong et al. created the PAFormer (Pyramid Attention Transformer) model, incorporating a performer module for overall aggregation and a SEAP (SEAttention and Atrous Spatial Pyramid Pooling) module for edge detection. Transformer models capture extensive dependencies and contextual details \cite{huang2022paformer}.

\subsection{Attention Based}

Angermanna et al.'s approach achieved a high 90\%  success rate in polyp detection on the CVC-ColonDB database, with an average processing time of just 0.039 seconds per frame \cite{angermann2016active}. Deeba et al. developed a polyp detection algorithm using computer assistance for colonoscopy and wireless capsule endoscopy. It outperformed existing methods regarding recall and F2 score  \cite{deeba2020computer}.
A new Fuzzy Attention module by Patel et al. improved polyp segmentation in colonoscopy images. It surpassed PraNet and achieved state-of-the-art accuracy on various datasets, outperforming recent methods like SA-Net, TransFuse, and Polyp-PVT \cite{patel2022fuzzynet}.
Du et al. proposed the ICGNet network for precise polyp segmentation in colonoscopic images.  ICGNet excels in learning ability and generalization compared to other methods, making it highly suitable for practical applications in colonoscopy \cite{du2022icgnet}.
Wei et al. introduced the Shallow Attention Network (SANet) for polyp segmentation. It outperformed previous methods and achieves a speed of about 72FPS \cite{wei2021shallow}.
Srivastava et al. proposed the Global Multi-Scale Residual Fusion Network (GMSRF-Net) with Cross Multi-Scale Attention (CMSA) and Multi-Scale Feature Selection (MSFS) modules to enhance polyp segmentation accuracy. The GMSRF-Net outperformed previous methods by 8.34\% and 10.31\% on the CVC-ClinicDB and Kvasir-SEG datasets \cite{srivastava2022gmsrf}.
Zhang et al. proposed SAMAug, a method that used segmentation masks generated by SAM to improve medical image segmentation models. The method demonstrated improved segmentation accuracy for the GlaS dataset in experiments \cite{li2024polyp}.
Wu et al. developed an ACL-Net framework for semi-supervised polyp segmentation using attribute contrastive learning. The method improved the accuracy of computer-aided therapy \cite{wu2023acl}. In a study, Joel et al. utilized The Guided Attention Residual Network (GAR-Net), which integrated residual blocks and attention mechanisms to identify and isolate polyps in colonoscopy images effectively \cite{raymann2022gar}. 
Weidong et al. proposed a model to improve colorectal polyp segmentation using a nonlinear activation-free method and a contextual attention module \cite{wu2023nonlinear}.

\subsection{U-Net Based}

Dinh et al. proposed the AG-CUResNeSt method, a unique design combining Attention Gates (AG) and Coupled UNets with a ResNeSt backbone \cite{sang2021ag}. 
Michael et al. employed an advanced deep neural network, Focus U-Net, which incorporates dual attention mechanisms and architectural enhancements like short-range skip connections and utilizes the Hybrid Focal loss for accurate polyp segmentation \cite{yeung2021focus}.
Trinh et al. have introduced a new approach to medical image segmentation by combining MetaFormer and UNet models. Their method features a Multi-scale Upsampling block to enhance texture and a Convformer block to improve local feature information \cite{trinh2023meta}.
Liao et al. introduced HarDNet-DFUS, a neural network architecture for medical image segmentation of diabetic foot ulcers and colonoscopy polyps.  The HarDNet-DFUS network outperformed HarDNet-MSEG in colonoscopy polyp segmentation \cite{liao2022hardnet}.
Shweta et al. developed an enhanced DeeplabV3+ model with a multi-level context attention mechanism to improve polyp segmentation. The model utilizes an ASPP module, ResNet101 backbone, and encoder-decoder architecture for effective polyp segmentation \cite{gangrade2024modified}.

\section{Proposed Method}
\label{method}

In this section, we outline the architecture of the proposed segmentation network and provide details of its component modules. We begin by briefly describing the architecture of the DeepLabV3+ \cite{rakshit2021multiclass} network, followed by a detailed explanation of our proposed architecture and the modules integrated within it. \textcolor{RoyalBlue}{Figure} \ref{DeepLabV3++}  illustrates the proposed architecture.
\subsection{Overview of DeeplabV3+}

For image segmentation tasks, DeepLabv3+ is a refined neural network model \cite{chen2018encoder}. In order to improve segmentation insights and better capture visual context, it builds upon the original DeepLabv3 by adding an encoder-decoder structure.

In DeepLabv3+, the encoder utilizes a backbone network such as ResNet or Xception incorporating Atrous Convolution at different layers to effectively capture multi-scale context while preserving spatial resolution \cite{rakshit2021multiclass,chen2018encoder}. Key elements of the model include an initial convolutional layer, an Atrous Spatial Pyramid Pooling (ASPP) module, and deep atrous convolution layers. The decoder improves the segmentation map by initially upscaling the encoded features and then merging them with corresponding lower-level features from the encoder. Refined convolution layers are then applied to enhance details and generate more precise object boundaries, resulting in superior segmentation outcomes. The final layer generates the segmentation map, typically followed by a softmax function for pixel-wise classification.
\subsection{Overview of proposed DeeplabV3++}

\begin{figure*}. 
\centering 		\includegraphics[width=\textwidth]{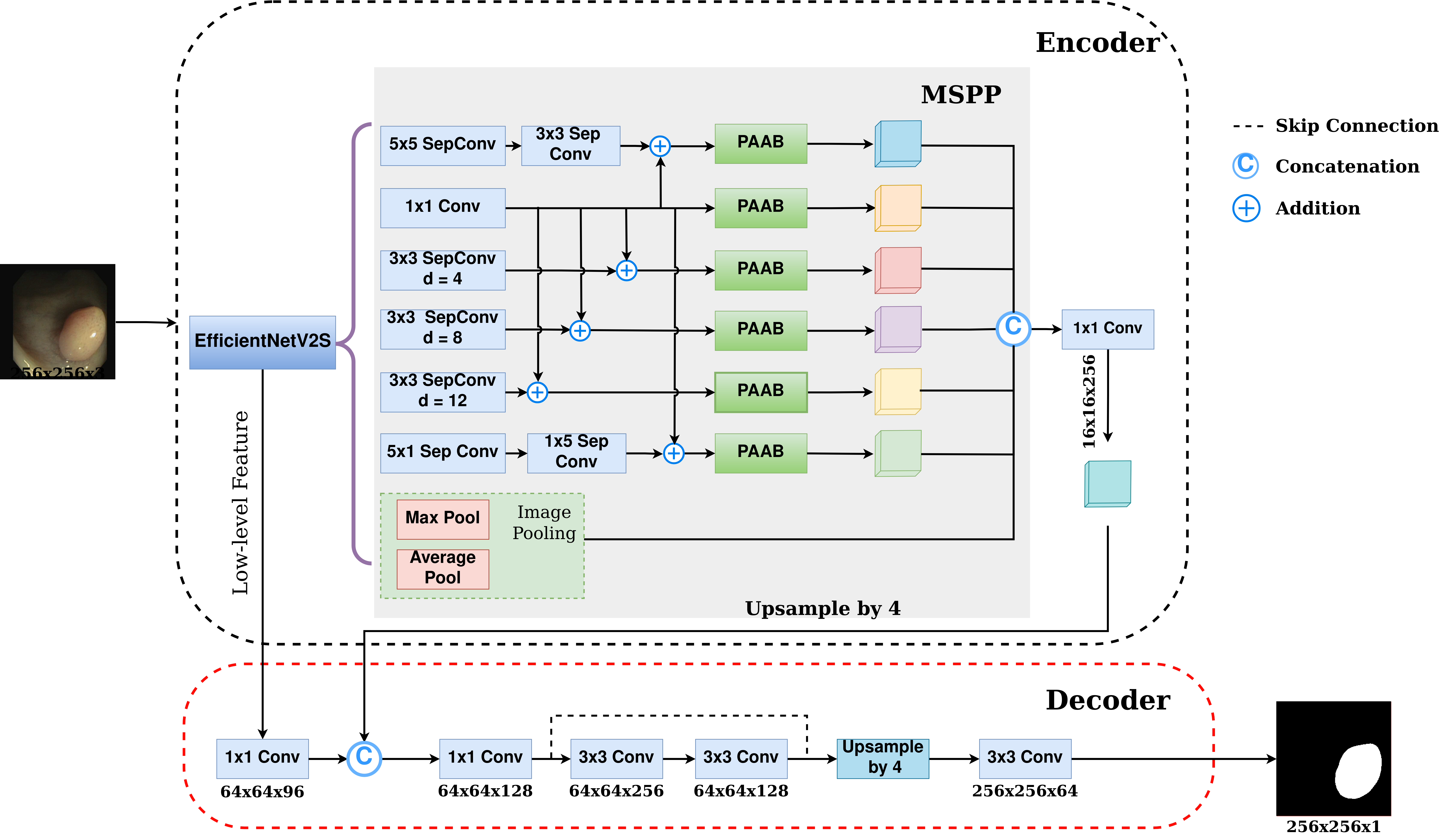}
\caption{Proposed DeepLabV3++ architecture}
\label{DeepLabV3++}
\end{figure*}

In our proposed DeepLabV3++ model, we have introduced Multi-Scale Pyramid Pooling (MSPP) with Parallel Attention Aggregation Block (PAAB) to achieve more robust polyp segmentation results from colonoscopy images. \textcolor{RoyalBlue}{Figure} \ref{DeepLabV3++} illustrates the architecture of our proposed model DeepLabV3++.

In the enhanced DeepLabV3++ model for polyp segmentation, we utilized EfficientNetV2S  as the encoder backbone to extract features from the input image. This is followed by the MSPP module, which incorporates convolution layers with varying dilation rates, skip connections, and pooling operations (max and average) to capture features at diverse scales. Skip connections in the model enhance gradient flow and preserve fine spatial details, improving segmentation accuracy and efficacy by including multi-scale features. Then, each feature undergoes enhancement through PAAB to emphasize crucial details. The decoder then begins by upsampling the encoded features and merging them with lower-level features from the encoder. It subsequently undergoes a series of convolution layers and skip connections to refine the segmentation map, followed by additional upsampling to align with the input resolution, resulting in precise and intricate polyp segmentation.
\subsection{Encoder}

The encoder of our proposed DeepLabv3++ model initiates with EfficientNetV2S as the backbone model for initial feature extraction. Subsequently, an MSPP module, including multiple convolution layers with skip connections and a pooling layer, is employed to encompass a range of scales for contextual information. The resulting outputs are channelled through PAAB to enhance significant features, assuring detailed feature extraction. The outputs from PAAB layers are then merged and passed through a \(1\times1\) convolution layer in order to reduce the number of channels and merge the multi-scale features efficiently. This procedure ensures that the resulting feature map is rich in detail and context for precise polyp segmentation. In this section, we will go over the encoder in detail.
\subsubsection{EfficientNetV2}

An enhanced version of EfficientNet, known as EfficientNetV2, has been specifically developed 6.8 times smaller in size to increase efficiency and speed up the training process \cite{tan2021efficientnetv2}. The primary advantages of EfficientNetV2 encompass reduced training time, decreased computational cost, and exceptional precision, rendering it well-suited for a wide array of computer vision assignments \cite{ghadami2024scalable,grd2024transfer,abioye2024performance}.

EfficientNetV2 integrated with the DeepLabv3+ model for semantic segmentation enhances accuracy and efficiency, providing superior results with reduced computational cost \cite{li2024sea,ullah2024concrete}. A recent study by Mati et al. proposed a lightweight segmentation model modified DeepLabV3+ with EfficientNetV2 as the backbone, which performed better than other existing CNN-based models \cite{ullah2024concrete}. Our proposed DeeplabV3++ model, which worked incredibly well for polyp segmentation, has pre-trained EfficientV2S as a feature extractor. Because of its refined feature extraction capabilities, which enhance segmentation accuracy, we have chosen EfficientV2S as our encoder backbone.
\subsubsection{Multi-Scale Pyramid Pooling}

In our proposed DeepLabV3++ model, the Multi-Scale Pyramid Pooling (MSPP) module, depicted in \textcolor{RoyalBlue}{Figure} \ref{DeepLabV3++}, replaces the Atrous Spatial Pyramid Pooling (ASPP) mechanism as an enhanced feature extraction technique that is intended to capture multi-scale contextual information more efficiently. This is accomplished by combining different convolutional and pooling layers, each adding something unique to the feature extraction procedure.

By splitting the typical convolution into depthwise and pointwise convolutions, the module's initial \(5\times5\) separable convolution layer and its subsequent \(3\times3\) separable convolution layer is added with \(1\times1\) convolution layer. As a result, more meaningful features can be captured at a bigger scale without affecting computational effectiveness. After that, dimension reduction and fine-grained feature extraction are carried out using a \(1\times1\) convolution layer. This phase reduces the number of channels in the feature map without losing significant information to preserve the fine features required for precisely identifying polyp borders. This layer is also added to all other convolutional layers present in MSPP. Integrating this \(1\times1\) convolution layer into other convolutional layers through a connection improves feature fusion by allowing the direct merging of multi-scale features. This makes combining detailed and contextual information easier, improving segmentation accuracy and preserving important spatial characteristics. By directly connecting lower-level features to higher layers, they enable the integration of multi-scale features. It also has multiple \(3\times3\) separable convolution layers with varying dilation rates (4, 8, and 12) to collect contextual information at different scales. The model can capture medium to large-scale context because dilated convolutions enlarge the receptive field without lowering spatial resolution \cite{chen2017rethinking}. This is necessary to comprehend polyp structure and its environments so that different-sized polyps can be recognized precisely. Additionally, the module uses \(5\times1\) separable convolutional layer and its subsequent \(1\times5\) separable convolution layer to capture directional information, emphasizing vertical and horizontal structures. This improves the model's capacity to identify directed and elongated features frequently present in the irregular forms of polyps. These directional properties help the model comprehend the polyps' structure, improving segmentation accuracy. It improves segmentation accuracy by combining contextual and detailed data. Furthermore, the module gathers global contextual information through average and max pooling methods. While average pooling captures the general contextual background, max pooling catches the most prominent elements. This combination improves the overall segmentation accuracy by effectively understanding visible and subtle features within the image. Afterwards, every output from these layers passes through the proposed PAAB module, which uses attention mechanisms to suppress irrelevant information and emphasize important ones. This makes sure the model pays attention to critical regions, such as textures and borders of polyp, which produces segmentation results that are more accurate \cite{liu2024image}.

\subsubsection{Parallel Attention Aggregation Block}

\begin{figure*}
	\centering
		\includegraphics[width=\textwidth]{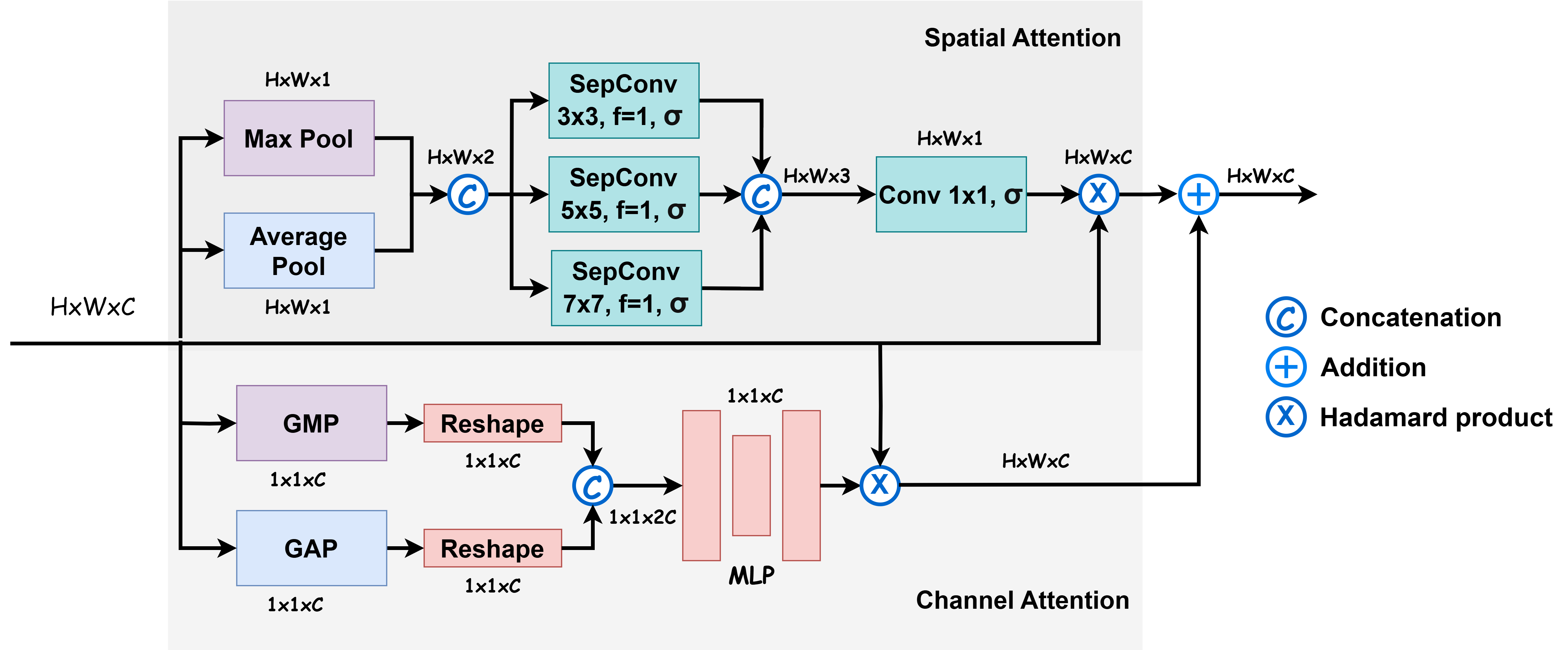}
	\caption{Architectural Composition of Parallel Attention Aggregation Block (PAAB)}
	\label{PAAB}
\end{figure*}

The proposed parallel attention aggregation block (PAAB), depicted in \textcolor{RoyalBlue}{Figure} \ref{PAAB}, improves the feature representation ability of deeplabv3++ by integrating spatial and channel attention mechanisms in parallel pathways. This dual-attention strategy effectively captures spatial relationships and channel-wise dependencies, enhancing feature representation. In the first branch, we initiate the aggregation of spatial information from the feature map by employing average and max pooling operations across the channels. This process generates average-pooled features and max-pooled features that enhance the modelling of channel inter-dependencies \cite{woo2018cbamconvolutionalblockattention}. The features are then concatenated and convolved by three separable convolution layers with kernel sizes of 3, 5, and 7, each using a sigmoid activation function. Unlike the standard convolutions, depth-wise convolutions with C=1 are significantly less computationally expensive. They perform operations independently on each channel rather than across all channels simultaneously. The outputs of these convolutions are concatenated and passed through a pointwise convolution, producing the 2D spatial attention map. These representations enhance the robustness and comprehensiveness of the model by capturing diverse aspects of spatial information. Given intermediate features \( F \in \mathbb{R}^{H \times W \times C} \) as input, the generated 2D spatial attention map \( M_s(F) \in \mathbb{R}^{H \times W} \) encodes the spatial locations to emphasize or suppress. The process can be described by the following equations:

\begin{equation}
\begin{aligned}
F_{\text{concat}} &= \text{Concat}(F_{c}^{\text{avg}}, F_{c}^{\text{max}}) \\
F_{dw3,5,7} &= \sigma(\text{DWConv2D}(F_{\text{concat}})) \\
F_{\text{merged}} &= \text{Concat}(F_{dw3}, F_{dw5}, F_{dw7})
\end{aligned}
\end{equation}

Apply a pointwise convolution to generate the spatial attention map:
\begin{equation}
M_s(F) = \sigma(\text{Conv2D}(F_{\text{merged}}, \text{k\_size}=1))
\label{eq:1}
\end{equation}
In the second branch, we aggregate channel information using both Global Average Pooling (GAP) and Global Max Pooling (GMP). The combined output then passes through an MLP with two hidden layers: the first layer uses ReLU activation, and the second layer uses sigmoid activation to produce the channel attention map \( M_c(F) \in \mathbb{R}^{1 \times 1 \times C} \).
\begin{align*}
F_{\text{concat}} = \text{Concat}(F_{\text{gap}}, F_{\text{gmp}}) \\
\end{align*} Apply two consecutive dense layers to generate the channel attention map: 
\begin{equation}
M_c(F) = \sigma(\text{Dense}(\text{ReLU}(\text{Dense}(F_{\text{concat}}))))
\label{eq:2}
\end{equation}
where \( \sigma \) denotes the sigmoid function. Finally, we use the Hadamard product to combine the input feature \( F \) with the spatial attention map \( M_s(F) \) and channel attention map \( M_c(F) \) to emphasize spatial features selectively. Afterwards, we perform element-wise addition on these two attention outputs. The refined feature is represented as:
\begin{equation}
F_{\text{refined}} = F \odot M_s(F) + F \odot M_c(F)
\label{eq:3}
\end{equation} 
where \( \odot \) denotes the Hadamard Product operation. The refined feature helps the model dynamically adapt to different polyp sizes. It captures fine details in small polyps and maintains contextual awareness in larger ones.

\subsection{Decoder}

The decoder refines and improves the segmentation output produced by the encoder in the proposed DeepLabv3++ model. By combining multi-scale feature fusion, comprehensive refining, and accurate upsampling, the decoder greatly improves the accuracy and dependability of polyp segmentation findings.

The encoded features are first upsampled by the decoder to bring them back to a higher resolution, which is necessary to accomplish thorough segmentation. In order to enable the model to produce a feature map corresponding to the spatial resolution of the lower-level data from the encoder, this upsampling is commonly accomplished via transposed convolutions or bilinear interpolation \cite{kundu2020attention}. This allows the decoder to efficiently combine contextual, high-level data with fine-grained features from the encoder that were maintained through skip connections. These upsampled features are concatenated with the matching lower-level features from the encoder by the decoder after the initial upsampling. The precise spatial information required for successful segmentation is preserved by the process of integrating multi-scale characteristics \cite{ronneberger2015u}. To ensure that both local and global contexts are considered in the final segmentation map, the concatenation phase uses the contextual information from the deeper levels of the encoder and the detailed information from the early layers. The characteristics are further refined, and the decoder enhances the segmentation quality by applying a sequence of \(1\times1\) convolution layer and \(3\times3\) convolution layers with skip connection after concatenation. To build more accurate borders around the segmented objects and minimise artefacts, convolutional layers are utilised \cite{long2015fully}.  In addition, the convolution layers with skip connections maintain fine spatial features and enhance gradient flow, which facilitates deep network training \cite{he2016deep}.

Then, the decoder executes an extra upsampling step to return the feature map to the initial input resolution. This guarantees that the size of the input image and the output segmentation map align. Finally, the decoder assembles the segmented image by mapping the upsampled features through a \(3\times3\) convolution layer. This stage highlights the regions of interest (polyps) in the input image by converting the refined feature map into a segmentation mask \cite{badrinarayanan2017segnet}.

\section{Experimental Analysis}
\label{exp}

This section covers the utilized datasets, pre-processing steps, chosen experimental parameters and evaluation metrics.

\subsection{Datasets}

We have chosen three popular benchmark datasets (CVC-ColonDB \cite{bernal2012towards}, CVC-ClinicDB \cite{bernal2015wm}, Kvasir-SEG \cite{jha2020kvasir} of colonoscopy images to demonstrate the capabilities of our proposed DeepLabV3++ model. We will cover the details of the datasets in this part. 
\subsubsection*{CVC-ColonDB}

The Universitat Autònoma de Barcelona's Computer Vision Center (CVC) has provided a collection of high-quality colonoscopy images and films called the CVC-ColonDB dataset \cite{bernal2012towards}. There are 380 images in the dataset. Every image has a \(574\times500\) pixel resolution. 15 brief videos of colonoscopies were used to extract the images. Medical specialists have thoroughly annotated every image and video frame in the CVC-ColonDB collection.

\subsubsection*{CVC-ClinicDB}

The open-access CVC-ClinicDB dataset includes 612 images of \(288\times384\) resolution \cite{bernal2015wm}. All the images were collected from 29 video sequences of colonoscopies. It was constructed in cooperation with a hospital in Barcelona, Spain \cite{li2024polyp}. Medical experts with extensive experience manually labelled each image in the dataset. where a wide variety of polyp sizes, forms, and looks are depicted in the images, offering a thorough visualization of actual clinical situations.

\subsubsection*{Kvasir-SEG}

The Kvasir-SEG is also an open-access dataset, and Vestre Viken Health Trust's gastroenterologists in Norway have confirmed the accuracy of the dataset \cite{jha2020kvasir}. It consists of 1000 polyp images, 1071 associated masks and bounding boxes, and their accompanying ground truth \cite{li2024polyp}. The pixels in the images vary in size from \( 332 \times 487 \) to \( 1920 \times 1072 \) \cite{krenzer2023real}. It is one of the largest gastrointestinal tract image datasets which is available to the public for polyp segmentation \cite{duc2022colonformer}.
\\
\\
The summary of the specifications for each of the three datasets we used for our study is shown in \textcolor{RoyalBlue}{Table} \protect\ref{tbl1}.

\subsection{Preprocessing and Data Augmentation}

To prepare our datasets for analysis, we have incorporated various pre-processing steps to ensure that the images are formatted appropriately for the training of our proposed model. We resized the input images to 256 $\times$ 256 and applied normalization to the input images. Resizing ensures all images have uniform dimensions, which facilitates compatibility with the model architecture, and normalization adjusts the pixel values to a consistent range for faster convergence and stable training. This procedure is essential for standardizing the input data and optimizing it to improve the robustness and adaptability of the model.
\begin{table*}[width=\textwidth,cols=5,pos=h]
\caption{Specifications of the datasets used in our research.}\label{tbl1}
\begin{tabular*}{\tblwidth}{@{} LLLLL@{} }
\toprule
\textbf{Dataset} & \textbf{No. of images} & \textbf{Image Size} & \textbf{Availability} & \textbf{Augmented Dataset Sample}\\
\midrule
CVC-ColonDB \cite{bernal2012towards} & 380 & \(574 \times 500\) & Public & 2672 \\
CVC-ClinicDB \cite{bernal2015wm} & 612 & \(288 \times 384\) & Public & 3900\\
Kvasir-SEG \cite{jha2020kvasir} & 1000 & Variable & Public & 4590\\
\bottomrule
\end{tabular*}
\end{table*}
Additionally, we incorporated several augmentation techniques specifically designed for biomedical image data to enhance the robustness and generalization of the model. These techniques include horizontal and vertical flips, random rotations, shift scale rotations, grid distortions and elastic transformations. The original images were also included in the training dataset.

\subsection{Exprimental Parameters}

To train the proposed model, the augmented dataset was split into three portions with an 80:10:10 ratio: 80\% of the images were allocated for training, 10\% for validation, and the remaining 10\% for testing. We initialized the pre-trained weights of the EfficientNetV2S \cite{tan2021efficientnetv2} architecture. For the CVC-ColonDB and CVC-ClinicDB datasets, the batch size was set to 4, and for the Kvasir-SEG dataset, it was set to 8. The initial learning rate was 0.0001 and was reduced by a factor of 0.1 after 15 epochs of no improvement. Early stopping was applied with a patience of 20 epochs, and the Adam optimizer was utilized to train the proposed model. Our system utilized a Tesla P100-PCIE GPU with 16 GB of RAM and was implemented with a TensorFlow backend. The proposed model comprises a total of 8.7 million trainable parameters. We employed a hybrid loss function to address the class imbalance problem and enhance the accuracy of segmented image boundaries. The hybrid loss function combines the Dice, binary cross-entropy loss (BCE), and focal loss. The hybrid loss function is defined as:
\begin{equation}
\mathcal{L}_{hybrid} = \alpha \cdot \mathcal{L}_{dice} + \beta \cdot \mathcal{L}_{bce} + \gamma \cdot \mathcal{L}_{focal}
\end{equation}

where
 \( \mathcal{L}_{dice} \) denotes the Dice loss,
 \( \mathcal{L}_{bce} \) denotes the binary cross-entropy loss,
 \( \mathcal{L}_{focal} \) denotes the focal loss and 
 \( \alpha, \beta, \gamma \) are coefficients to balance the contributions of each loss function. Here, the binary cross-entropy loss, dice loss and focal loss is computed as follows:
\begin{equation}
\begin{aligned}
\mathcal{L}_{BCE} &= -\frac{1}{N} \sum_{d=1}^{N} y_{o,d} \log(p_{o,d}) \\
                  &= -( y_k \log(p) + (1 - y_k) \log(1 - p))
\end{aligned}
\end{equation}
where, \( y_k \) is the true label and \( p \) is the predicted probability.

\begin{equation}
\mathcal{L}_{Dice} = 1 - \frac{2 \sum_{i=1}^{T} c_i d_i + \epsilon}{\sum_{i=1}^{T} c_i^2 + \sum_{i=1}^{T} d_i^2 + \epsilon}
\end{equation}
where, \( c_i \) and \( d_i \) represent the predicted and ground truth values respectively for pixel \( i \), \( T \) is the total number of pixels and \( \epsilon \) is a small constant added for numerical stability.

\begin{equation}
\mathcal{L}_{focal} = \sum_{k=1}^{C} \beta_k \left( - (1 - x_{k})^\gamma \cdot \log(x_{k}) \right)
\end{equation}
where, \( C \) is the number of classes, \( x_k \) is the predicted probability for class \( k \), \( \beta_k \) is the balancing factor for class \( k \) and \( \gamma \) is the focusing parameter.

The summary of our training parameters for each of the used datasets is presented in \textcolor{RoyalBlue}{Table} \protect\ref{tbl5}.

\begin{table*}[width=\textwidth,cols=6,pos=h]
\caption{The distinct characteristics of the learning environment for the models in each dataset.}\label{tbl5}
\begin{tabular*}{\tblwidth}{@{} LLLLLL@{} }
\toprule
\textbf{Dataset} & \textbf{Loss} & \textbf{LR} & \textbf{Epoch} & \textbf{Batch Size} & \textbf{Input Size}\\
\midrule
CVC-ColonDB \cite{bernal2012towards} & BCE + FOCAL + DC & \(1e^{-4}\) & 150 & 4 & \(256 \times 256\)\\
CVC-ClinicDB \cite{bernal2015wm} & BCE + FOCAL + DC & \(1e^{-4}\) & 150 & 8 & \(256 \times 256\) \\
Kvasir-SEG \cite{jha2020kvasir} & BCE + DC & \(1e^{-4}\) & 150 & 8 & \(256 \times 256\)\\
\bottomrule
\end{tabular*}
\end{table*}

\subsection{Evaluation Metrices}

We employed various evaluation metrics to analyze the performance of our polyp segmentation model. Dice Coefficient, Precision, Recall, Accuracy, and Mean Intersection over Union (mIoU) are the metrics that we employed in this study to measure our evaluation outcomes. The following is a description of the metrics equations:

\begin{equation}
\text{Dice Coefficient} =\frac{2|A \cap A'|}{|A| + |A'|}
\end{equation}

\begin{equation}
\text{Precision} = \frac{\text{TP}}{\text{TP} + \text{FP}}
\end{equation}

\begin{equation}
\text{Recall} = \frac{\text{TP}}{\text{TP} + \text{FN}}
\end{equation}

\begin{equation}
\text{Accuracy} = \frac{\text{TP} + \text{TN}}{\text{TP} + \text{TN} + \text{FP} + \text{FN}}
\end{equation}

\begin{equation}
\text{mIoU} = \frac{|A \cap A'|}{|A \cup A'|}
\end{equation}

Here, The areas of the ground truth and predicted masks are denoted by $A$ and $A'$, respectively and \text{FN} represents the number of false negatives, \text{TP} represents the number of true positives, \text{TN} represents the number of true negatives, \text{FP} represents the number of false positives.

\section{Discussion}
\label{discussion}

In the field of medical imaging, accurate segmentation of areas of interest (such as polyp) is crucial for clinical evaluation. Our proposed DeepLabv3++ model, incorporating MSPP and PAAB, has displayed exceptional performance on a range of polyp segmentation datasets. The MSPP component outperforms the traditional ASPP method by offering extensive multi-scale feature extraction, effectively capturing features of various sizes and shapes. Moreover, the PAAB improves segmentation precision by utilizing attention mechanisms that prioritize important regions such as polyp boundaries and textures. This combination enables our model to achieve more accurate and consistent segmentation results than other existing SOTA models, confirming its strong proficiency in polyp segmentation.
\subsection{Comparison with the existing SOTA methods}

We have evaluated the performance of our proposed DeepLabV3++ on three publicly available polyp datasets. To benchmark its efficacy, we conducted comparative analyses against several state-of-the-art segmentation methods, including Unet3+ \cite{hatamizadeh2021unetrtransformers3dmedical}, MultiresUnet \cite{ibtehaz2020multiresunet}, UnetR \cite{hatamizadeh2021unetrtransformers3dmedical}, Polyp-PVT \cite{dong2021polyp}, DeepLabV3+ \cite{rakshit2021multiclass}, and Duck-Net \cite{dumitru2023using}. These models were reproduced using open-source codes and their original implementations to provide fair comparisons. Moreover, we also standardized the augmentation techniques, data splitting ratio, and loss function for all the methods under comparison to prevent any performance bias. \textcolor{RoyalBlue} {Table} \protect\ref{tbl2} showcased the performance comparison of the DeepLabV3++ with these existing methods. The segmentation results are displayed in terms of Accuracy, Dice Coefficient (DICE), Mean Intersection Over Union (mIoU), Precision, and Recall. From \textcolor{RoyalBlue} {Table} \protect\ref{tbl2} we can see that our proposed model performs better than a number of different existing models in each of the evaluation metrics. For the CVC-ColonDB dataset, our model achieves an impressive accuracy of 98.59\%, an mIoU of 94.74\%, a DICE score of 96.20\%, Precision of 97.85\% and Recall of 94.29\%. These results outperform traditional models such as MultiResUNet (DICE: 85.10\%, mIoU: 78.36\%, Precision: 94.49\%, Recall: 91.28\%), UNet3+ (DICE: 88.07\%, mIoU: 83.58\%, Precision: 94.63\%, Recall: 91.86\%), as well as advanced models like Polyp-PVT (DICE: 93.16\%, mIoU: 87.91\%, Precision: 93.30\%, Recall: 92.04\%) and DUCK-Net (DICE: 93.53\%, mIoU: 87.85\%, Precision: 93.14\%, Recall: 93.92\%), highlighting the resilience and efficacy of our methodology in capturing intricate details and contextual features. Similarly our proposed DeeplabV3++ model obtains mIoU values of 95.63\%, 93.78\%, Precision of 96.35\%, 95.22\%, Recall of 94.42\%, 94.58\%, as well as DICE scores of 96.54\%, 96.08\% and Accuracy of 98.78\%, 99.12\% in the CVC-ClinicDB, kvasir-SEG datasets, respectively. For all of these datasets our model outperformed all other existing model's DICE and mIoU scores. Additionally, our model outperformed the majority of the other models in terms of recall, accuracy, and precision also.

\begin{figure*}
	\centering
		\includegraphics[scale=.36]{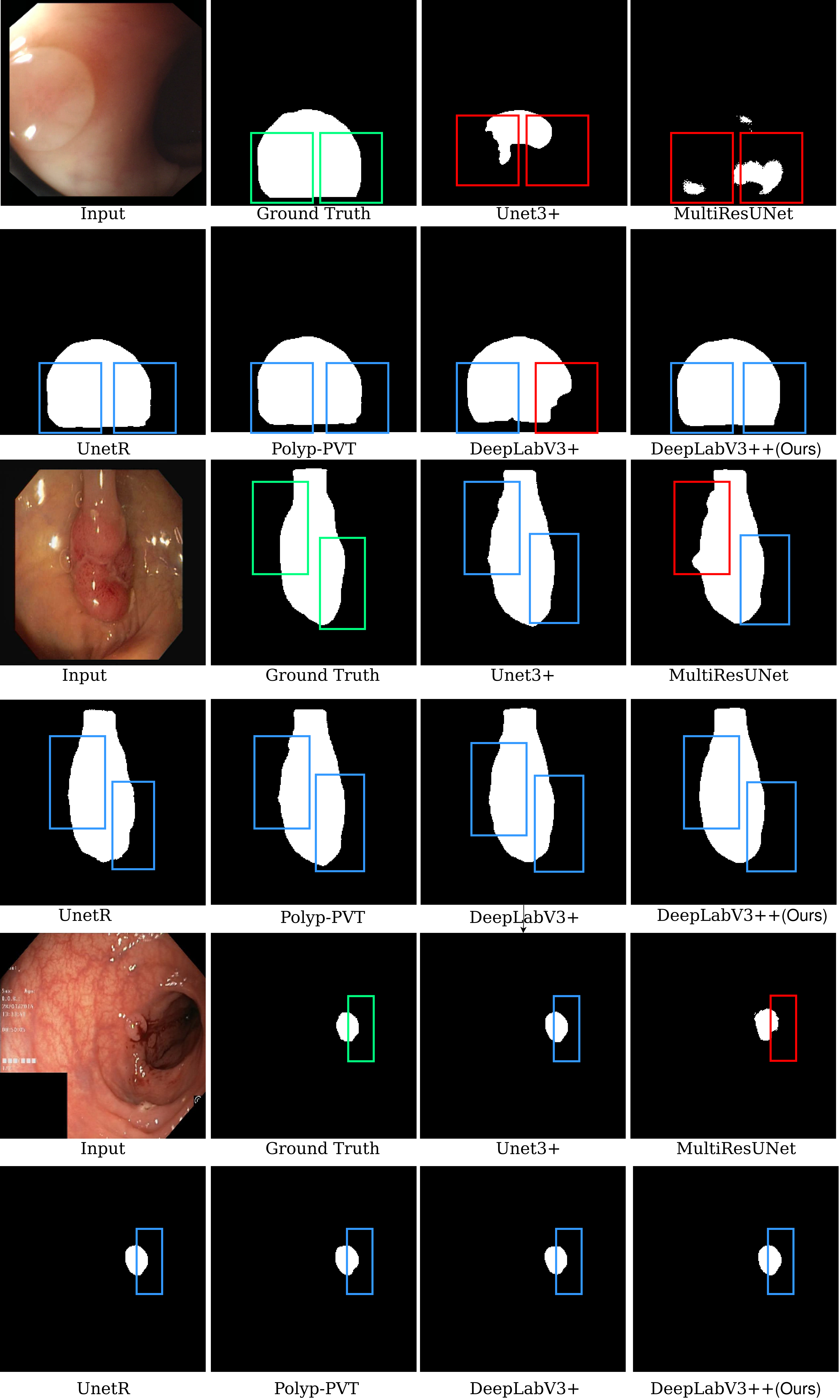}
	\caption{Qualitative Comparison of Segmentation Results Using Unet3+, MultiResUnet, UnetR, Polyp-PVT, DeepLabV3+, and proposed DeepLabV3++ Models on CVC-ColonDB, CVC-ClinicDB and Kvasir-SEG datasets. Here, green, blue, and red boxes represent exemplary ROI and acceptable and unsatisfactory results.}
	\label{FIG:3}
\end{figure*}

\begin{table*}[width=\textwidth,cols=8,pos=h]
\caption{Comparison of segmentation results between the proposed and conventional methods across all datasets employed. Bold numbers indicate the best performance for each corresponding indicator. "-" denotes absence of a backbone used in the network and also not mentioned.}\label{tbl2}

\begin{tabular*}{\tblwidth}{@{} LLLLLLLL@{} }
\toprule
\textbf{Dataset} & Method & Backbone & Accuracy & mIoU & DICE & Precision & Recall\\
\midrule
CVC-ColonDB & MultiResUNet \cite{ibtehaz2020multiresunet} & - & 92.74 & 78.36 & 85.10 & 94.49 & 91.28\\
 & UNet3+ \cite{huang2020unet3fullscaleconnected} & - & 94.22 & 83.58 & 88.07 & 94.63 & 91.86 \\
 & U-NetR \cite{hatamizadeh2021unetrtransformers3dmedical} & Vision Transformer & 96.71 & 89.26 & 93.55 & 94.18 & 93.40 \\
 & Polyp-PVT \cite{dong2021polyp} & Pyramid vision transformer & 96.11 & 87.91 & 93.16 & 93.30 & 92.04 \\
 & DUCK-Net \cite{dumitru2023using} & - & \textbf{99.29} & 87.85 & 93.53 & 93.14 & 93.92 \\
 & EMCAD \cite{rahman2024emcad} & - & - & - & 92.31 & - & - \\
 & UGCANet \cite{hung2023ugcanet} & - & - & 74.90 & 82.70 & - & - \\
 & MEGANet \cite{bui2024meganet} & Res2Net-50 & - & 71.40 & 79.30 & - & - \\
 & Deeplabv3+ \cite{rakshit2021multiclass} & ResNet50 & 95.39 & 90.56 & 92.40 & 94.78 & 92.37\\
 & \textbf{DeepLabV3++} (Ours) & EfficientNetV2S & 98.59 &  \textbf{94.74} &  \textbf{96.20} &  \textbf{97.85} &  \textbf{94.29}\\
 CVC-ClinicDB & MultiResUnet \cite{ibtehaz2020multiresunet} & - & 93.12 & 88.17 & 90.09 & 95.67 & 92.82 \\
 & UNet3+ \cite{huang2020unet3fullscaleconnected} & - & 94.40 & 90.15 & 94.32 & 96.12 & 94.22 \\
 & UnetR \cite{hatamizadeh2021unetrtransformers3dmedical} & Vision Transformer & 96.37 & 90.33 & 93.17 & 95.89 & 92.10\\
 & Polyp-PVT \cite{dong2021polyp} & Pyramid vision transformer & 95.11 & 89.31 & 94.49 & 95.47 & 93.39 \\
 & Duck-Net \cite{dumitru2023using} & - & 99.07 & 90.09 & 94.78 & 94.68 &  \textbf{94.89} \\
 & EMCAD \cite{rahman2024emcad} & - & - & - & 95.21 & - & - \\
 & UGCANet \cite{hung2023ugcanet} & - & - & 90.70 & 95.00 & - & - \\
 & MEGANet \cite{bui2024meganet} & Res2Net-50 & - & 89.40 & 93.80 & - & - \\
 & DOLG-NeXt \cite{ahmed2023dolg} & Conv-NeXt &  \textbf{99.14} &  91.80 &  95.10 &  96.30 & 94.88 \\
 & Deeplabv3+ \cite{rakshit2021multiclass} & ResNet50 & 95.26 & 90.39 & 93.24 & 94.65 & 89.72\\
 & \textbf{DeepLabV3++} (Ours) & EfficientNetV2S & 98.78 &  \textbf{95.63} &  \textbf{96.54}
&  \textbf{96.35} & 94.42\\
Kvasir-SEG & MultiResUnet \cite{ibtehaz2020multiresunet} & - & 95.12 & 89.17 & 93.79 & 93.67 & 90.82 \\
 & UNet3+ \cite{huang2020unet3fullscaleconnected} & - & 96.66 & 89.05 & 94.62 & 94.42 & 93.39 \\
 & UnetR \cite{hatamizadeh2021unetrtransformers3dmedical} & Vision Transformer & 96.97 & 92.70 & 94.54 & 94.79 & 89.10\\
 & Polyp-PVT \cite{dong2021polyp} & Pyramid vision transformer & 95.91 & 89.11 & 93.49 & 93.63 & 91.19 \\
 & Duck-Net \cite{dumitru2023using} & - & 98.42 & 90.51 & 95.02 &  \textbf{96.28} & 93.79 \\
 & EMCAD \cite{rahman2024emcad} & - & - & - & 92.75 & - & - \\
 & UGCANet \cite{hung2023ugcanet} & - & - & 87.4 & 92.6 & - & -\\
 & MEGANet \cite{bui2024meganet} & Res2Net-50 & - & 86.3 & 91.3 & - & - \\
 & DeepLabv3+ \cite{rakshit2021multiclass} & ResNet50 & 96.69 & 90.19 & 92.33 & 94.18 & 92.49\\
 & \textbf{DeepLabV3++} (Ours) & EfficientNetV2S &  \textbf{99.12} &  \textbf{93.78} &  \textbf{96.08} &  95.22 &  \textbf{94.58}\\
\bottomrule
\end{tabular*}
\end{table*}
The metrics demonstrate how effectively our model can adapt to different polyp datasets, consistently achieving high levels of precision and recall. Our model shows better segmentation performance when compared to other state-of-the-art techniques such as DOLG-NeXt and Duck-Net, especially when it comes to identifying the complex polyp borders and textures. By incorporating EfficientNetV2S as model backbone and a multi-scale feature extraction method, our model is better equipped to handle the diverse and intricate structures present in biomedical images. In addition to achieving high performance metrics, our approach guarantees consistent and dependable segmentation results across a range of polyp datasets. A few of the predicted masks from the three datasets we used for the qualitative comparison are shown in \textcolor{RoyalBlue} {Figure} \ref{FIG:3}. Observations indicate that our proposed model consistently produces segmentation masks that closely match the actual boundaries, showcasing its proficiency in effectively segmenting uniform regions with varying sizes and textures. Particularly noteworthy is its expertise in capturing the intricate details of polyp edges, which are typically complex, high-contrast, and blurred.

\subsection{Ablation study}

In this work, we perform a comprehensive evaluation of our DeepLabV3++ model through ablation studies. We employ EfficientNetV2S \cite{tan2021efficientnetv2} as the backbone for our DeepLabV3++ architecture. To assess the impact of various components within our model, we conduct the ablation studies on CVC-ColonDB, CVC-ClinicDB, and Kvasir-SEG datasets. The detailed results of these ablation studies are presented in \textcolor{RoyalBlue} {Table} \ref{tbl3}.
\begin{table*}[width=\textwidth,cols=8,pos=h]
\caption{Ablation Study Results}\label{tbl3}
\begin{tabular*}{\tblwidth}{@{} LLLLLLLL@{} }
\toprule
Dataset & Model & mIoU & DICE & Precision & Recall & Parameters\\
\midrule
CVC-ColonDB & Baseline  & 90.56  & 92.40 & 94.78  & 92.37 & 17,795,425 \\
& DeepLabV3++ w/o MSPP & 88.26 & 91.13 & 92.41 & 83.17 & 6,259,497 \\
& DeepLabV3++ w/o PAAB in MSPP  & 94.11 & 95.03 & 96.66 & 93.67 &  8,792,937 \\
&  \textbf{Proposed DeepLabV3++} & 94.74  & 96.20 & 97.85  & 94.29 & 8,793,832 \\
\midrule
 CVC-ClinicDB & Baseline & 90.39  & 93.24 & 94.65 & 89.72 & 17,795,425 \\
& DeepLabV3++ w/o MSPP & 91.89  & 94.34 & 94.05 & 92.67 & 6,259,497 \\
& DeepLabV3++ w/o PAAB in MSPP & 95.83 & 96.14 & 95.52 & 94.60 &  8,792,937 \\
& \textbf{ Proposed DeepLabV3++}  & 95.63  & 96.54 & 96.35  & 94.42 & 8,793,832  \\
\midrule
Kvasir-SEG & Baseline & 90.19  & 92.33  & 94.18 & 92.49 & 17,795,425 \\
& DeepLabV3++ w/o MSPP & 89.39   & 92.29 & 92.81 & 91.10 & 6,259,497 \\
& DeepLabV3++ w/o PAAB in MSPP  & 92.71 & 95.67 & 94.12 & 93.87 &  8,792,937 \\
&  \textbf{Proposed DeepLabV3++} & 93.78  & 96.08 & 95.22 & 94.58 & 8,793,832 \\
\bottomrule
\end{tabular*}
\end{table*}
A baseline DeepLabV3+ with  Xception backbone was utilized to benchmark the performance of the datasets. The results confirm that the successive addition of each component contributes to improved segmentation performance. From \textcolor{RoyalBlue} {Table} \ref{tbl3} we can see that, the final model achieves impressive DICE scores (96.20\%, 96.54\%, and 96.08\%) and mIoU scores (94.74\%, 95.63\%, and 93.78\%), compared to the baseline DeepLabV3+ model (DICE: 92.40\%,  93.24\%, 92.33\%  and  mIoU: 90.56\%,  90.39\%,  90.19\%) on the CVC-ColonDB, CVC-ClinicDB, and Kvasir-SEG datasets, respectively. 

As shown in \textcolor{RoyalBlue} {Table} \ref{tbl3}, incorporating the MSPP module within DeepLabV3++'s encoder leads to a significant improvement in performance on the CVC-ColonDB, CVC-ClinicDB and Kvasir-SEG dataset. Compared to the proposed model architecture without MSPP, the model with MSPP achieves 5.07\%, 2.2\% and 3.79\% higher DICE score and 6.48\%, 3.74\% and 4.39\% higher mIoU score, respectively. The incorporation of the PAAB within the MSPP module also contributes to the enhanced performance of the DeepLabV3++ model. The proposed model with the PAAB module inside MSPP performs slightly better (DICE: 1.17\%, 0.4\% and 0.41\% higher) than model without PAAB in MSPP on CVC-ColonDB, CVC-ClinicDB and Kvasir-SEG datasets, respectively.  Additionally, the proposed DeepLabV3++ model achieves a 2.02\% reduction in parameters compared to the baseline DeepLabV3+ architecture, resulting in improved inference time.
\subsection{Enhancing Polyp Segmentation with DeepLabV3++}

The existing DeepLabV3+ model utilizes a ResNet-50 backbone for feature extraction and incorporates atrous spatial pyramid pooling (ASPP) to capture context at different scales. Although this design is efficient at segmenting complicated images, it struggles to handle diverse polyp sizes and capture fine details. In contrast, our proposed DeepLabV3++ model enhances the backbone with EfficientNetV2S, providing a better trade-off between accuracy and efficiency. The model includes multiple \(3\times3\) separable convolutions with varying dilation rates to improve contextual information capture. Additionally, a \(1\times1\) convolution layer is added to reduce dimensions and extract detailed features by adding it to other convolutional layers in Multi-Scale Pyramid Pooling (MSPP) module, while max and average pooling methods refine global contextual insight. The model also integrates PAAB modules for attention focusing on important areas and concealing irrelevant details. This comprehensive approach results in superior performance in polyp segmentation tasks, achieving higher accuracy, mIoU, and DICE scores compared to the baseline DeepLabV3+ model.

The proposed DeepLabV3++ model has significantly smaller number of trainable parameters than the existing DeepLabV3+ model with ResNet-50 as backbone. It has 8,793,832 million parameters, while the baseline model has more than twice as many, or 17,795,425 million trainable parameters. This is another strength of our model since it reduces training time and improves computational efficiency. The segmentation performance of our proposed DeepLabV3++ model with EfficientNetV2S as backbone, greatly outperforms the majority of prior SOTA methods found in literature review. Specifically, our model excels in handling intricate boundary areas of polyp's, addressing shape irregularities, and precisely classifying unclear images, showcasing improved robustness and accuracy compared to existing models.

\begin{figure*}
	\centering
		\includegraphics[scale=.40]{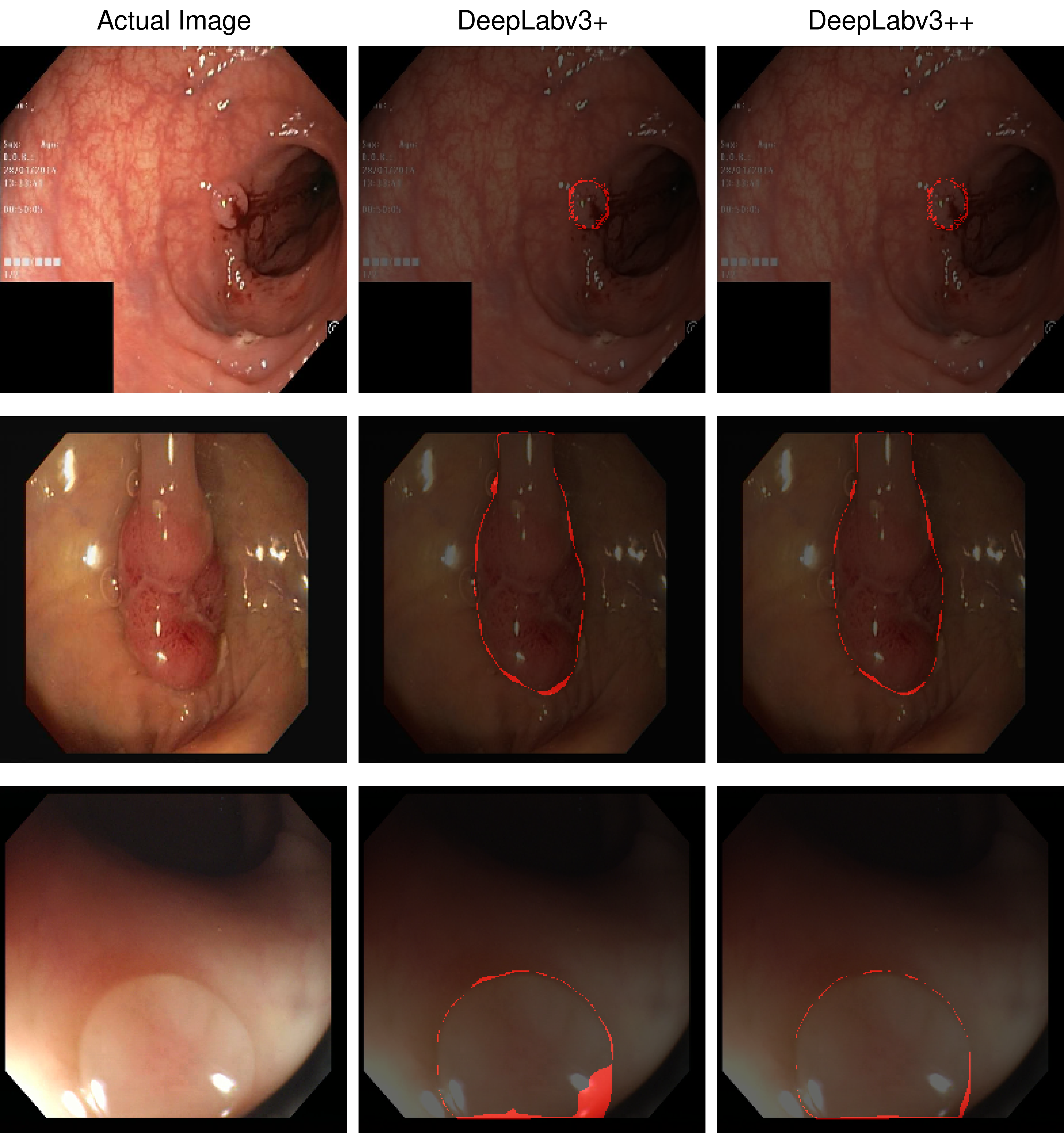}
	\caption{Segmentation error comparison of Deeplabv3+ and proposed Deeplabv3++. Here, the highlighted area represents the X-OR difference between the actual and predicted masks.}
	\label{error}
\end{figure*}

\textcolor{RoyalBlue}{Figure} \protect\ref{error} illustrates the comparison of segmentation error between DeeplabV3+ and proposed DeeplabV3++. In comparison to models such as UNet3+, MultiResUNet, DeeplabV3+ as well as advanced transformer-based models like UnetR and Polyp-PVT, our model exhibits reduced noise and fewer errors in segmentation. This leads to clearer and more precise masks, guaranteeing a high degree of adaptability across diverse datasets and types of polyps. In \textcolor{RoyalBlue}{Figure} \protect\ref{error} we can see the x-or difference (highlighted areas) between the actual masks and predicted masks of the models individually. Here differences are shown using an x-or operation, where matched regions in the segmentation mask are denoted by 0 (unhighlighted), whereas mismatched regions are denoted by 1 (highlighted). Less highlighted areas imply greater accuracy, and the highlighted areas show segmentation errors. The figure makes it clear that, in comparison to Deeplabv3+, the proposed DeepLabV3++ model displays noticeably less highlighted areas. This shows how much better our suggested model is in segmenting polyp areas, as evidenced by the far fewer segmentation errors it achieves. The decrease in highlighted regions reflects to DeepLabV3++'s resilience in managing intricate boundary regions, uneven shapes, and fuzzy images, all of which greatly improve segmentation quality.

\section{Conclusion\label{concl}}

In this paper, we introduced a novel segmentation model called DeepLabV3++ to precisely segment the regions of interest (ROIs) in colonoscopy images.
The DeepLabV3++ model, using EfficientNetV2S as its backbone, not only achieves better segmentation results but also shows efficiency with significantly fewer trainable parameters and fewer segmentation errors compared to existing models. The model combines MSPP and PAAB to address the challenges encountered in diverse polyp datasets. The MSPP enhances traditional ASPP through comprehensive multi-scale feature extraction. The incorporation of PAAB and separable convolutions further boosts our model's segmentation performance while maintaining computational efficiency. It is evident that DeepLabV3++ outperforms other state-of-the-art models on three benchmark public datasets of colonoscopy images. Across the CVC-ColonDB, CVC-ClinicDB, and Kvasir-SEG datasets, this proposed model achieves mIoU values of 94.74\%, 95.63\%, and 93.78\%, as well as DICE scores of 96.20\%, 96.54\%, and 96.08\%, respectively.

To further enhance DeepLabV3++'s efficiency, we intend to explore advanced optimization techniques like quantization methods and efficient architectural modifications. Additionally, our goal is to expand the application of DeepLa-bV3++ to other medical image segmentation tasks beyond polyp segmentation. Finally, we know that incorporation of a transformer encoder architecture shows promise in capturing longer-range dependencies and enhancing global context within images. We plan to evaluate the effectiveness of this approach in future research.






\printcredits


\bibliography{cas-refs}

\begin{thebibliography}{86}
\providecommand{\natexlab}[1]{#1}
\providecommand{\url}[1]{\texttt{#1}}
\expandafter\ifx\csname urlstyle\endcsname\relax
  \providecommand{\doi}[1]{doi: #1}\else
  \providecommand{\doi}{doi: \begingroup \urlstyle{rm}\Url}\fi

\bibitem[Paschke et~al.(2018)Paschke, Jafarov, Staib, Kreuser, Maulbecker-Armstrong, Roitman, Holm, Harris, Link, and Kornmann]{paschke2018colon}
Stephan Paschke, Sakhavat Jafarov, Ludger Staib, Ernst-Dietrich Kreuser, Catharina Maulbecker-Armstrong, Marc Roitman, Torbj{\"o}rn Holm, Curtis~C Harris, Karl-Heinrich Link, and Marko Kornmann.
\newblock Are colon and rectal cancer two different tumor entities? a proposal to abandon the term colorectal cancer.
\newblock \emph{International journal of molecular sciences}, 19\penalty0 (9):\penalty0 2577, 2018.

\bibitem[Labianca et~al.(2010)Labianca, Beretta, Kildani, Milesi, Merlin, Mosconi, Pessi, Prochilo, Quadri, Gatta, {de Braud}, and Wils]{LABIANCA2010106}
Roberto Labianca, Giordano~D. Beretta, Basem Kildani, Laura Milesi, Federica Merlin, Stefania Mosconi, M.~Adelaide Pessi, Tiziana Prochilo, Antonello Quadri, Gemma Gatta, Filippo {de Braud}, and Jacques Wils.
\newblock Colon cancer.
\newblock \emph{Critical Reviews in Oncology/Hematology}, 74\penalty0 (2):\penalty0 106--133, 2010.
\newblock ISSN 1040-8428.
\newblock \doi{https://doi.org/10.1016/j.critrevonc.2010.01.010}.
\newblock URL \url{https://www.sciencedirect.com/science/article/pii/S1040842810000119}.

\bibitem[Bretthauer et~al.(2022)Bretthauer, L{\o}berg, Wieszczy, Kalager, Emilsson, Garborg, Rupinski, Dekker, Spaander, Bugajski, et~al.]{bretthauer2022effect}
Michael Bretthauer, Magnus L{\o}berg, Paulina Wieszczy, Mette Kalager, Louise Emilsson, Kjetil Garborg, Maciej Rupinski, Evelien Dekker, Manon Spaander, Marek Bugajski, et~al.
\newblock Effect of colonoscopy screening on risks of colorectal cancer and related death.
\newblock \emph{New England Journal of Medicine}, 387\penalty0 (17):\penalty0 1547--1556, 2022.

\bibitem[Xi and Xu(2021)]{xi2021global}
Yue Xi and Pengfei Xu.
\newblock Global colorectal cancer burden in 2020 and projections to 2040.
\newblock \emph{Translational oncology}, 14\penalty0 (10):\penalty0 101174, 2021.

\bibitem[Sung et~al.(2021)Sung, Ferlay, Siegel, Laversanne, Soerjomataram, Jemal, and Bray]{sung2021global}
Hyuna Sung, Jacques Ferlay, Rebecca~L Siegel, Mathieu Laversanne, Isabelle Soerjomataram, Ahmedin Jemal, and Freddie Bray.
\newblock Global cancer statistics 2020: Globocan estimates of incidence and mortality worldwide for 36 cancers in 185 countries.
\newblock \emph{CA: a cancer journal for clinicians}, 71\penalty0 (3):\penalty0 209--249, 2021.

\bibitem[Amersi et~al.(2005)Amersi, Agustin, and Ko]{amersi2005colorectal}
Farin Amersi, Michelle Agustin, and Clifford~Y Ko.
\newblock Colorectal cancer: epidemiology, risk factors, and health services.
\newblock \emph{Clinics in colon and rectal surgery}, 18\penalty0 (03):\penalty0 133--140, 2005.

\bibitem[Ahmed(2020)]{ahmed2020colon}
Monjur Ahmed.
\newblock Colon cancer: a clinician’s perspective in 2019.
\newblock \emph{Gastroenterology research}, 13\penalty0 (1):\penalty0 1, 2020.

\bibitem[Allen(1995)]{allen1995molecular}
John~I Allen.
\newblock Molecular biology of colon polyps and colon cancer.
\newblock In \emph{Seminars in surgical oncology}, volume~11, pages 399--405. Wiley Online Library, 1995.

\bibitem[Cappell(2005)]{cappell2005colonic}
Mitchell~S Cappell.
\newblock From colonic polyps to colon cancer: pathophysiology, clinical presentation, and diagnosis.
\newblock \emph{Clinics in laboratory medicine}, 25\penalty0 (1):\penalty0 135--177, 2005.

\bibitem[Sullivan et~al.(2022)Sullivan, Noujaim, and Roper]{sullivan2022cause}
Brian~A Sullivan, Michael Noujaim, and Jatin Roper.
\newblock Cause, epidemiology, and histology of polyps and pathways to colorectal cancer.
\newblock \emph{Gastrointestinal Endoscopy Clinics}, 32\penalty0 (2):\penalty0 177--194, 2022.

\bibitem[Gangrade et~al.(2024)Gangrade, Sharma, Sharma, and Singh]{gangrade2024modified}
Shweta Gangrade, Prakash~Chandra Sharma, Akhilesh~Kumar Sharma, and Yadvendra~Pratap Singh.
\newblock Modified deeplabv3+ with multi-level context attention mechanism for colonoscopy polyp segmentation.
\newblock \emph{Computers in Biology and Medicine}, page 108096, 2024.

\bibitem[Giovannucci(2002)]{giovannucci2002modifiable}
Edward Giovannucci.
\newblock Modifiable risk factors for colon cancer.
\newblock \emph{Gastroenterology Clinics}, 31\penalty0 (4):\penalty0 925--943, 2002.

\bibitem[Ananthakrishnan and Xavier(2020)]{ANANTHAKRISHNAN202016}
Ashwin~N. Ananthakrishnan and Ramnik~J. Xavier.
\newblock 3 - gastrointestinal diseases.
\newblock In Edward~T. Ryan, David~R. Hill, Tom Solomon, Naomi~E. Aronson, and Timothy~P. Endy, editors, \emph{Hunter's Tropical Medicine and Emerging Infectious Diseases (Tenth Edition)}, pages 16--26. Elsevier, London, tenth edition edition, 2020.
\newblock ISBN 978-0-323-55512-8.
\newblock \doi{https://doi.org/10.1016/B978-0-323-55512-8.00003-X}.
\newblock URL \url{https://www.sciencedirect.com/science/article/pii/B978032355512800003X}.

\bibitem[Pacal et~al.(2020)Pacal, Karaboga, Basturk, Akay, and Nalbantoglu]{pacal2020comprehensive}
Ishak Pacal, Dervis Karaboga, Alper Basturk, Bahriye Akay, and Ufuk Nalbantoglu.
\newblock A comprehensive review of deep learning in colon cancer.
\newblock \emph{Computers in Biology and Medicine}, 126:\penalty0 104003, 2020.

\bibitem[Tasnim et~al.(2021)Tasnim, Chakraborty, Shamrat, Chowdhury, Nuha, Karim, Zahir, and Billah]{tasnim2021deep}
Zarrin Tasnim, Sovon Chakraborty, FM~Javed~Mehedi Shamrat, Ali~Newaz Chowdhury, Humaira~Alam Nuha, Asif Karim, Sabrina~Binte Zahir, and Md~Masum Billah.
\newblock Deep learning predictive model for colon cancer patient using cnn-based classification.
\newblock \emph{International Journal of Advanced Computer Science and Applications}, 12\penalty0 (8):\penalty0 687--696, 2021.

\bibitem[Tharwat et~al.(2022)Tharwat, Sakr, El-Sappagh, Soliman, Kwak, and Elmogy]{tharwat2022colon}
Mai Tharwat, Nehal~A Sakr, Shaker El-Sappagh, Hassan Soliman, Kyung-Sup Kwak, and Mohammed Elmogy.
\newblock Colon cancer diagnosis based on machine learning and deep learning: modalities and analysis techniques.
\newblock \emph{Sensors}, 22\penalty0 (23):\penalty0 9250, 2022.

\bibitem[Thambawita et~al.(2021)Thambawita, Hicks, Halvorsen, and Riegler]{thambawita2021divergentnets}
Vajira Thambawita, Steven~A Hicks, P{\aa}l Halvorsen, and Michael~A Riegler.
\newblock Divergentnets: Medical image segmentation by network ensemble.
\newblock \emph{arXiv preprint arXiv:2107.00283}, 2021.

\bibitem[Tomar et~al.(2022)Tomar, Shergill, Rieders, Bagci, and Jha]{tomar2022transresu}
Nikhil~Kumar Tomar, Annie Shergill, Brandon Rieders, Ulas Bagci, and Debesh Jha.
\newblock Transresu-net: Transformer based resu-net for real-time colonoscopy polyp segmentation.
\newblock \emph{arXiv preprint arXiv:2206.08985}, 2022.

\bibitem[Tomar et~al.(2023)Tomar, Jha, and Bagci]{tomar2023dilatedsegnet}
Nikhil~Kumar Tomar, Debesh Jha, and Ulas Bagci.
\newblock Dilatedsegnet: A deep dilated segmentation network for polyp segmentation.
\newblock In \emph{International Conference on Multimedia Modeling}, pages 334--344. Springer, 2023.

\bibitem[Huang et~al.(2021)Huang, Wu, and Lin]{huang2021hardnet}
Chien-Hsiang Huang, Hung-Yu Wu, and Youn-Long Lin.
\newblock Hardnet-mseg: A simple encoder-decoder polyp segmentation neural network that achieves over 0.9 mean dice and 86 fps.
\newblock \emph{arXiv preprint arXiv:2101.07172}, 2021.

\bibitem[Huang et~al.(2018)Huang, Xiao, Chang, Tsai, and Liu]{huang2018automatic}
Cheng-Hsien Huang, Wei-Ting Xiao, Li-Jen Chang, Wei-Ta Tsai, and Wei-Min Liu.
\newblock Automatic tissue segmentation by deep learning: from colorectal polyps in colonoscopy to abdominal organs in ct exam.
\newblock In \emph{2018 IEEE Visual Communications and Image Processing (VCIP)}, pages 1--4. IEEE, 2018.

\bibitem[Lan et~al.(2023)Lan, Chen, and Jin]{lan2023dri}
Xiaoke Lan, Honghuan Chen, and Wenbing Jin.
\newblock Dri-net: segmentation of polyp in colonoscopy images using dense residual-inception network.
\newblock \emph{Frontiers in Physiology}, 14:\penalty0 1290820, 2023.

\bibitem[Ronneberger et~al.(2015)Ronneberger, Fischer, and Brox]{ronneberger2015u}
Olaf Ronneberger, Philipp Fischer, and Thomas Brox.
\newblock U-net: Convolutional networks for biomedical image segmentation.
\newblock In \emph{Medical image computing and computer-assisted intervention--MICCAI 2015: 18th international conference, Munich, Germany, October 5-9, 2015, proceedings, part III 18}, pages 234--241. Springer, 2015.

\bibitem[Punn and Agarwal(2022)]{punn2022modality}
Narinder~Singh Punn and Sonali Agarwal.
\newblock Modality specific u-net variants for biomedical image segmentation: a survey.
\newblock \emph{Artificial Intelligence Review}, 55\penalty0 (7):\penalty0 5845--5889, 2022.

\bibitem[Siddique et~al.(2021)Siddique, Paheding, Elkin, and Devabhaktuni]{9446143}
Nahian Siddique, Sidike Paheding, Colin~P. Elkin, and Vijay Devabhaktuni.
\newblock U-net and its variants for medical image segmentation: A review of theory and applications.
\newblock \emph{IEEE Access}, 9:\penalty0 82031--82057, 2021.
\newblock \doi{10.1109/ACCESS.2021.3086020}.

\bibitem[Safarov and Whangbo(2021)]{safarov2021denseunet}
Sirojbek Safarov and Taeg~Keun Whangbo.
\newblock A-denseunet: Adaptive densely connected unet for polyp segmentation in colonoscopy images with atrous convolution.
\newblock \emph{Sensors}, 21\penalty0 (4):\penalty0 1441, 2021.

\bibitem[Ngoc~Lan et~al.(2021)Ngoc~Lan, An, Hang, Long, Trung, Thuy, and Sang]{ngoc2021neounet}
Phan Ngoc~Lan, Nguyen~Sy An, Dao~Viet Hang, Dao~Van Long, Tran~Quang Trung, Nguyen~Thi Thuy, and Dinh~Viet Sang.
\newblock Neounet: Towards accurate colon polyp segmentation and neoplasm detection.
\newblock In \emph{Advances in Visual Computing: 16th International Symposium, ISVC 2021, Virtual Event, October 4-6, 2021, Proceedings, Part II}, pages 15--28. Springer, 2021.

\bibitem[Raymann and Rajalakshmi(2022)]{raymann2022gar}
Joel Raymann and Ratnavel Rajalakshmi.
\newblock Gar-net: Guided attention residual network for polyp segmentation from colonoscopy video frames.
\newblock \emph{Diagnostics}, 13\penalty0 (1):\penalty0 123, 2022.

\bibitem[Xu et~al.(2021)Xu, Zhao, Yu, Zhang, Bian, Wang, Ge, and Qian]{xu2021real}
Jianwei Xu, Ran Zhao, Yizhou Yu, Qingwei Zhang, Xianzhang Bian, Jun Wang, Zhizheng Ge, and Dahong Qian.
\newblock Real-time automatic polyp detection in colonoscopy using feature enhancement module and spatiotemporal similarity correlation unit.
\newblock \emph{Biomedical Signal Processing and Control}, 66:\penalty0 102503, 2021.

\bibitem[Lee et~al.(2020)Lee, Jeong, Song, Ha, Lee, Koo, Yang, Kim, and Byeon]{lee2020real}
Ji~Young Lee, Jinhoon Jeong, Eun~Mi Song, Chunae Ha, Hyo~Jeong Lee, Ja~Eun Koo, Dong-Hoon Yang, Namkug Kim, and Jeong-Sik Byeon.
\newblock Real-time detection of colon polyps during colonoscopy using deep learning: systematic validation with four independent datasets.
\newblock \emph{Scientific reports}, 10\penalty0 (1):\penalty0 8379, 2020.

\bibitem[Luo et~al.(2021)Luo, Zhang, Liu, Lai, Liu, Wang, Xing, Huang, Li, Li, et~al.]{luo2021artificial}
Yuchen Luo, Yi~Zhang, Ming Liu, Yihong Lai, Panpan Liu, Zhen Wang, Tongyin Xing, Ying Huang, Yue Li, Aiming Li, et~al.
\newblock Artificial intelligence-assisted colonoscopy for detection of colon polyps: a prospective, randomized cohort study.
\newblock \emph{Journal of Gastrointestinal Surgery}, 25\penalty0 (8):\penalty0 2011--2018, 2021.

\bibitem[Nur-A-Alam et~al.(2024)Nur-A-Alam, Uddin, Manu, Rahman, and Nasir]{nur2024automatic}
Md~Nur-A-Alam, Khandaker Mohammad~Mohi Uddin, MMR Manu, Md~Mahbubur Rahman, and Mostofa~Kamal Nasir.
\newblock An automatic system to detect colorectal polyp using hybrid fused method from colonoscopy images.
\newblock \emph{Intelligent Systems with Applications}, page 200342, 2024.

\bibitem[Huang et~al.(2022)Huang, Xie, Qing, Wang, Liu, and Sun]{huang2022paformer}
Zhiyong Huang, Fang Xie, WenCheng Qing, MengYao Wang, Man Liu, and Daming Sun.
\newblock Paformer: Pyramid attention transformer for polyp segmentation.
\newblock 2022.

\bibitem[Hosseinzadeh~Kassani et~al.(2020)Hosseinzadeh~Kassani, Hosseinzadeh~Kassani, Wesolowski, Schneider, and Deters]{hosseinzadeh2020automatic}
Sara Hosseinzadeh~Kassani, Peyman Hosseinzadeh~Kassani, Michal~J Wesolowski, Kevin~A Schneider, and Ralph Deters.
\newblock Automatic polyp segmentation using convolutional neural networks.
\newblock In \emph{Advances in Artificial Intelligence: 33rd Canadian Conference on Artificial Intelligence, Canadian AI 2020, Ottawa, ON, Canada, May 13--15, 2020, Proceedings 33}, pages 290--301. Springer, 2020.

\bibitem[Chen et~al.(2021)Chen, Lu, Yu, Luo, Adeli, Wang, Lu, Yuille, and Zhou]{chen2021transunet}
Jieneng Chen, Yongyi Lu, Qihang Yu, Xiangde Luo, Ehsan Adeli, Yan Wang, Le~Lu, Alan~L Yuille, and Yuyin Zhou.
\newblock Transunet: Transformers make strong encoders for medical image segmentation.
\newblock \emph{arXiv preprint arXiv:2102.04306}, 2021.

\bibitem[Qadir et~al.(2021)Qadir, Shin, Solhusvik, Bergsland, Aabakken, and Balasingham]{qadir2021toward}
Hemin~Ali Qadir, Younghak Shin, Johannes Solhusvik, Jacob Bergsland, Lars Aabakken, and Ilangko Balasingham.
\newblock Toward real-time polyp detection using fully cnns for 2d gaussian shapes prediction.
\newblock \emph{Medical Image Analysis}, 68:\penalty0 101897, 2021.

\bibitem[Nisha and Gopi(2022)]{nisha2022colorectal}
JS~Nisha and Varun~Palakuzhiyil Gopi.
\newblock Colorectal polyp detection in colonoscopy videos using image enhancement and discrete orthonormal stockwell transform.
\newblock \emph{S{\=a}dhan{\=a}}, 47\penalty0 (4):\penalty0 234, 2022.

\bibitem[Dash et~al.(2023)Dash, Dash, Padhy, Das, Mishra, and Paikaray]{dash2023identification}
AB~Dash, S~Dash, S~Padhy, RK~Das, B~Mishra, and BK~Paikaray.
\newblock Identification of polyp from colonoscopy images by deep belief network based polyp detector integration model.
\newblock \emph{EAI Endorsed Transactions on Pervasive Health and Technology}, 9, 2023.

\bibitem[Huang et~al.()Huang, Wu, and Lin]{huang2101hardnet}
Chien-Hsiang Huang, Hung-Yu Wu, and Youn-Long Lin.
\newblock Hardnet-mseg: A simple encoder-decoder polyp segmentation neural network that achieves over 0.9 mean dice and 86 fps. arxiv 2021.
\newblock \emph{arXiv preprint arXiv:2101.07172}.

\bibitem[Zhang et~al.(2021)Zhang, Liu, and Hu]{zhang2021transfuse}
Yundong Zhang, Huiye Liu, and Qiang Hu.
\newblock Transfuse: Fusing transformers and cnns for medical image segmentation.
\newblock In \emph{Medical Image Computing and Computer Assisted Intervention--MICCAI 2021: 24th International Conference, Strasbourg, France, September 27--October 1, 2021, Proceedings, Part I 24}, pages 14--24. Springer, 2021.

\bibitem[Shen et~al.(2021)Shen, Fu, Lin, and Zheng]{shen2021cotr}
Zhiqiang Shen, Rongda Fu, Chaonan Lin, and Shaohua Zheng.
\newblock Cotr: Convolution in transformer network for end to end polyp detection.
\newblock In \emph{2021 7th International Conference on Computer and Communications (ICCC)}, pages 1757--1761. IEEE, 2021.

\bibitem[Wang et~al.(2022)Wang, Huang, Tang, Meng, Su, and Song]{wang2022stepwise}
Jinfeng Wang, Qiming Huang, Feilong Tang, Jia Meng, Jionglong Su, and Sifan Song.
\newblock Stepwise feature fusion: Local guides global.
\newblock In \emph{International Conference on Medical Image Computing and Computer-Assisted Intervention}, pages 110--120. Springer, 2022.

\bibitem[Nguyen et~al.(2022)Nguyen, Bui, Van~Nguyen, Nguyen, and Van~Pham]{nguyen2022lapformer}
Mai Nguyen, Tung~Thanh Bui, Quan Van~Nguyen, Thanh~Tung Nguyen, and Toan Van~Pham.
\newblock Lapformer: A light and accurate polyp segmentation transformer.
\newblock \emph{arXiv preprint arXiv:2210.04393}, 2022.

\bibitem[Elkarazle et~al.(2023)Elkarazle, Raman, Then, and Chua]{elkarazle2023improved}
Khaled Elkarazle, Valliappan Raman, Patrick Then, and Caslon Chua.
\newblock Improved colorectal polyp segmentation using enhanced ma-net and modified mix-vit transformer.
\newblock \emph{IEEE Access}, 11:\penalty0 69295--69309, 2023.

\bibitem[Angermann et~al.(2016)Angermann, Histace, and Romain]{angermann2016active}
Quentin Angermann, Aymeric Histace, and Olivier Romain.
\newblock Active learning for real time detection of polyps in videocolonoscopy.
\newblock \emph{Procedia Computer Science}, 90:\penalty0 182--187, 2016.

\bibitem[Deeba et~al.(2020)Deeba, Bui, and Wahid]{deeba2020computer}
Farah Deeba, Francis~M Bui, and Khan~A Wahid.
\newblock Computer-aided polyp detection based on image enhancement and saliency-based selection.
\newblock \emph{Biomedical signal processing and control}, 55:\penalty0 101530, 2020.

\bibitem[Patel et~al.(2022)Patel, Li, and Wang]{patel2022fuzzynet}
Krushi~Bharatbhai Patel, Fengjun Li, and Guanghui Wang.
\newblock Fuzzynet: A fuzzy attention module for polyp segmentation.
\newblock In \emph{NeurIPS'22 Workshop on All Things Attention: Bridging Different Perspectives on Attention}, 2022.

\bibitem[Du et~al.(2022)Du, Xu, and Ma]{du2022icgnet}
Xiuquan Du, Xuebin Xu, and Kunpeng Ma.
\newblock Icgnet: Integration context-based reverse-contour guidance network for polyp segmentation.
\newblock In \emph{IJCAI}, pages 877--883, 2022.

\bibitem[Wei et~al.(2021)Wei, Hu, Zhang, Li, Zhou, and Cui]{wei2021shallow}
Jun Wei, Yiwen Hu, Ruimao Zhang, Zhen Li, S~Kevin Zhou, and Shuguang Cui.
\newblock Shallow attention network for polyp segmentation.
\newblock In \emph{Medical Image Computing and Computer Assisted Intervention--MICCAI 2021: 24th International Conference, Strasbourg, France, September 27--October 1, 2021, Proceedings, Part I 24}, pages 699--708. Springer, 2021.

\bibitem[Srivastava et~al.(2022)Srivastava, Chanda, Jha, Pal, and Ali]{srivastava2022gmsrf}
Abhishek Srivastava, Sukalpa Chanda, Debesh Jha, Umapada Pal, and Sharib Ali.
\newblock Gmsrf-net: An improved generalizability with global multi-scale residual fusion network for polyp segmentation.
\newblock In \emph{2022 26th International Conference on Pattern Recognition (ICPR)}, pages 4321--4327. IEEE, 2022.

\bibitem[Li et~al.(2024{\natexlab{a}})Li, Hu, and Yang]{li2024polyp}
Yuheng Li, Mingzhe Hu, and Xiaofeng Yang.
\newblock Polyp-sam: Transfer sam for polyp segmentation.
\newblock In \emph{Medical Imaging 2024: Computer-Aided Diagnosis}, volume 12927, pages 759--765. SPIE, 2024{\natexlab{a}}.

\bibitem[Wu et~al.(2023{\natexlab{a}})Wu, Xie, Lin, and Guo]{wu2023acl}
Huisi Wu, Wende Xie, Jingyin Lin, and Xinrong Guo.
\newblock Acl-net: semi-supervised polyp segmentation via affinity contrastive learning.
\newblock In \emph{Proceedings of the AAAI Conference on Artificial Intelligence}, volume~37, pages 2812--2820, 2023{\natexlab{a}}.

\bibitem[Wu et~al.(2023{\natexlab{b}})Wu, Fan, Fan, and Wen]{wu2023nonlinear}
Weidong Wu, Hongbo Fan, Yu~Fan, and Jian Wen.
\newblock Nonlinear activation-free contextual attention network for polyp segmentation.
\newblock \emph{Information}, 14\penalty0 (7):\penalty0 362, 2023{\natexlab{b}}.

\bibitem[Sang et~al.(2021)Sang, Chung, Lan, Hang, Van~Long, and Thuy]{sang2021ag}
Dinh~Viet Sang, Tran~Quang Chung, Phan~Ngoc Lan, Dao~Viet Hang, Dao Van~Long, and Nguyen~Thi Thuy.
\newblock Ag-curesnest: A novel method for colon polyp segmentation.
\newblock \emph{arXiv preprint arXiv:2105.00402}, 2021.

\bibitem[Yeung et~al.(2021)Yeung, Sala, Sch{\"o}nlieb, and Rundo]{yeung2021focus}
Michael Yeung, Evis Sala, Carola-Bibiane Sch{\"o}nlieb, and Leonardo Rundo.
\newblock Focus u-net: A novel dual attention-gated cnn for polyp segmentation during colonoscopy.
\newblock \emph{Computers in biology and medicine}, 137:\penalty0 104815, 2021.

\bibitem[Trinh(2023)]{trinh2023meta}
Quoc-Huy Trinh.
\newblock Meta-polyp: a baseline for efficient polyp segmentation.
\newblock In \emph{2023 IEEE 36th International Symposium on Computer-Based Medical Systems (CBMS)}, pages 742--747. IEEE, 2023.

\bibitem[Liao et~al.(2022)Liao, Yang, Lo, Lai, Shen, and Lin]{liao2022hardnet}
Ting-Yu Liao, Ching-Hui Yang, Yu-Wen Lo, Kuan-Ying Lai, Po-Huai Shen, and Youn-Long Lin.
\newblock Hardnet-dfus: An enhanced harmonically-connected network for diabetic foot ulcer image segmentation and colonoscopy polyp segmentation.
\newblock \emph{arXiv preprint arXiv:2209.07313}, 2022.

\bibitem[Rakshit(2021)]{rakshit2021multiclass}
Soumik Rakshit.
\newblock Multiclass semantic segmentation using deeplabv3+, 2021.

\bibitem[Chen et~al.(2018)Chen, Zhu, Papandreou, Schroff, and Adam]{chen2018encoder}
Liang-Chieh Chen, Yukun Zhu, George Papandreou, Florian Schroff, and Hartwig Adam.
\newblock Encoder-decoder with atrous separable convolution for semantic image segmentation.
\newblock In \emph{Proceedings of the European conference on computer vision (ECCV)}, pages 801--818, 2018.

\bibitem[Tan and Le(2021)]{tan2021efficientnetv2}
Mingxing Tan and Quoc Le.
\newblock Efficientnetv2: Smaller models and faster training.
\newblock In \emph{International conference on machine learning}, pages 10096--10106. PMLR, 2021.

\bibitem[Ghadami et~al.(2024)Ghadami, Rezvanian, and Shakuri]{ghadami2024scalable}
Omid Ghadami, Alireza Rezvanian, and Saeed Shakuri.
\newblock Scalable real-time emotion recognition using efficientnetv2 and resolution scaling.
\newblock In \emph{2024 10th International Conference on Web Research (ICWR)}, pages 7--12. IEEE, 2024.

\bibitem[Grd et~al.(2024)Grd, Tomi{\v{c}}i{\'c}, and Bar{\v{c}}i{\'c}]{grd2024transfer}
Petra Grd, Igor Tomi{\v{c}}i{\'c}, and Ena Bar{\v{c}}i{\'c}.
\newblock Transfer learning with efficientnetv2s for automatic face shape classification.
\newblock \emph{Journal of Universal Computer Science (JUCS)}, 30\penalty0 (2), 2024.

\bibitem[Abioye et~al.(2024)Abioye, Evwiekpaefe, and Awujoola]{abioye2024performance}
Oluwasegun~A Abioye, Abraham~E Evwiekpaefe, and Awujoola~J Awujoola.
\newblock Performance evaluation of efficientnetv2 models on the classification of histopathological benign breast cancer images.
\newblock \emph{Science Journal of University of Zakho}, 12\penalty0 (2):\penalty0 208--214, 2024.

\bibitem[Li et~al.(2024{\natexlab{b}})Li, Wang, Wu, Sun, Shi, and Ma]{li2024sea}
Sheng Li, Min Wang, Jia Wu, Shuo Sun, Minghang Shi, and Rui Ma.
\newblock Sea ice detection network for icebreakers in polar environments with attention-based deeplabv3+ architecture.
\newblock In \emph{Journal of Physics: Conference Series}, volume 2718, page 012062. IOP Publishing, 2024{\natexlab{b}}.

\bibitem[Ullah et~al.(2024)Ullah, Mir, Husain, Shahid, and Ahmad]{ullah2024concrete}
Mati Ullah, Junaid Mir, Syed~Sameed Husain, Muhammad Laiq Ur~Rahman Shahid, and Afaq Ahmad.
\newblock Concrete forensic analysis using deep learning-based coarse aggregate segmentation.
\newblock \emph{Automation in Construction}, 162:\penalty0 105372, 2024.

\bibitem[Chen et~al.(2017)Chen, Papandreou, Schroff, and Adam]{chen2017rethinking}
Liang-Chieh Chen, George Papandreou, Florian Schroff, and Hartwig Adam.
\newblock Rethinking atrous convolution for semantic image segmentation.
\newblock \emph{arXiv preprint arXiv:1706.05587}, 2017.

\bibitem[Liu et~al.(2024)Liu, Bai, Wang, Li, Li, and Lv]{liu2024image}
Yanyan Liu, Xiaotian Bai, Jiafei Wang, Guoning Li, Jin Li, and Zengming Lv.
\newblock Image semantic segmentation approach based on deeplabv3 plus network with an attention mechanism.
\newblock \emph{Engineering Applications of Artificial Intelligence}, 127:\penalty0 107260, 2024.

\bibitem[Woo et~al.(2018)Woo, Park, Lee, and Kweon]{woo2018cbamconvolutionalblockattention}
Sanghyun Woo, Jongchan Park, Joon-Young Lee, and In~So Kweon.
\newblock Cbam: Convolutional block attention module, 2018.
\newblock URL \url{https://arxiv.org/abs/1807.06521}.

\bibitem[Kundu et~al.(2020)Kundu, Mostafa, Sridhar, and Sundaresan]{kundu2020attention}
Souvik Kundu, Hesham Mostafa, Sharath~Nittur Sridhar, and Sairam Sundaresan.
\newblock Attention-based image upsampling.
\newblock \emph{arXiv preprint arXiv:2012.09904}, 2020.

\bibitem[Long et~al.(2015)Long, Shelhamer, and Darrell]{long2015fully}
Jonathan Long, Evan Shelhamer, and Trevor Darrell.
\newblock Fully convolutional networks for semantic segmentation.
\newblock In \emph{Proceedings of the IEEE conference on computer vision and pattern recognition}, pages 3431--3440, 2015.

\bibitem[He et~al.(2016)He, Zhang, Ren, and Sun]{he2016deep}
Kaiming He, Xiangyu Zhang, Shaoqing Ren, and Jian Sun.
\newblock Deep residual learning for image recognition.
\newblock In \emph{Proceedings of the IEEE conference on computer vision and pattern recognition}, pages 770--778, 2016.

\bibitem[Badrinarayanan et~al.(2017)Badrinarayanan, Kendall, and Cipolla]{badrinarayanan2017segnet}
Vijay Badrinarayanan, Alex Kendall, and Roberto Cipolla.
\newblock Segnet: A deep convolutional encoder-decoder architecture for image segmentation.
\newblock \emph{IEEE transactions on pattern analysis and machine intelligence}, 39\penalty0 (12):\penalty0 2481--2495, 2017.

\bibitem[Bernal et~al.(2012)Bernal, S{\'a}nchez, and Vilarino]{bernal2012towards}
Jorge Bernal, Javier S{\'a}nchez, and Fernando Vilarino.
\newblock Towards automatic polyp detection with a polyp appearance model.
\newblock \emph{Pattern Recognition}, 45\penalty0 (9):\penalty0 3166--3182, 2012.

\bibitem[Bernal et~al.(2015)Bernal, S{\'a}nchez, Fern{\'a}ndez-Esparrach, Gil, Rodr{\'\i}guez, and Vilari{\~n}o]{bernal2015wm}
Jorge Bernal, F~Javier S{\'a}nchez, Gloria Fern{\'a}ndez-Esparrach, Debora Gil, Cristina Rodr{\'\i}guez, and Fernando Vilari{\~n}o.
\newblock Wm-dova maps for accurate polyp highlighting in colonoscopy: Validation vs. saliency maps from physicians.
\newblock \emph{Computerized medical imaging and graphics}, 43:\penalty0 99--111, 2015.

\bibitem[Jha et~al.(2020)Jha, Smedsrud, Riegler, Halvorsen, De~Lange, Johansen, and Johansen]{jha2020kvasir}
Debesh Jha, Pia~H Smedsrud, Michael~A Riegler, P{\aa}l Halvorsen, Thomas De~Lange, Dag Johansen, and H{\aa}vard~D Johansen.
\newblock Kvasir-seg: A segmented polyp dataset.
\newblock In \emph{MultiMedia Modeling: 26th International Conference, MMM 2020, Daejeon, South Korea, January 5--8, 2020, Proceedings, Part II 26}, pages 451--462. Springer, 2020.

\bibitem[Krenzer et~al.(2023)Krenzer, Banck, Makowski, Hekalo, Fitting, Troya, Sudarevic, Zoller, Hann, and Puppe]{krenzer2023real}
Adrian Krenzer, Michael Banck, Kevin Makowski, Amar Hekalo, Daniel Fitting, Joel Troya, Boban Sudarevic, Wolfgang~G Zoller, Alexander Hann, and Frank Puppe.
\newblock A real-time polyp-detection system with clinical application in colonoscopy using deep convolutional neural networks.
\newblock \emph{Journal of imaging}, 9\penalty0 (2):\penalty0 26, 2023.

\bibitem[Duc et~al.(2022)Duc, Oanh, Thuy, Triet, and Dinh]{duc2022colonformer}
Nguyen~Thanh Duc, Nguyen~Thi Oanh, Nguyen~Thi Thuy, Tran~Minh Triet, and Viet~Sang Dinh.
\newblock Colonformer: An efficient transformer based method for colon polyp segmentation.
\newblock \emph{IEEE Access}, 10:\penalty0 80575--80586, 2022.

\bibitem[Hatamizadeh et~al.(2021)Hatamizadeh, Tang, Nath, Yang, Myronenko, Landman, Roth, and Xu]{hatamizadeh2021unetrtransformers3dmedical}
Ali Hatamizadeh, Yucheng Tang, Vishwesh Nath, Dong Yang, Andriy Myronenko, Bennett Landman, Holger Roth, and Daguang Xu.
\newblock Unetr: Transformers for 3d medical image segmentation, 2021.
\newblock URL \url{https://arxiv.org/abs/2103.10504}.

\bibitem[Ibtehaz and Rahman(2020)]{ibtehaz2020multiresunet}
Nabil Ibtehaz and M~Sohel Rahman.
\newblock Multiresunet: Rethinking the u-net architecture for multimodal biomedical image segmentation.
\newblock \emph{Neural networks}, 121:\penalty0 74--87, 2020.

\bibitem[Dong et~al.(2021)Dong, Wang, Fan, Li, Fu, and Shao]{dong2021polyp}
Bo~Dong, Wenhai Wang, Deng-Ping Fan, Jinpeng Li, Huazhu Fu, and Ling Shao.
\newblock Polyp-pvt: Polyp segmentation with pyramid vision transformers.
\newblock \emph{arXiv preprint arXiv:2108.06932}, 2021.

\bibitem[Dumitru et~al.(2023)Dumitru, Peteleaza, and Craciun]{dumitru2023using}
Razvan-Gabriel Dumitru, Darius Peteleaza, and Catalin Craciun.
\newblock Using duck-net for polyp image segmentation.
\newblock \emph{Scientific reports}, 13\penalty0 (1):\penalty0 9803, 2023.

\bibitem[Huang et~al.(2020)Huang, Lin, Tong, Hu, Zhang, Iwamoto, Han, Chen, and Wu]{huang2020unet3fullscaleconnected}
Huimin Huang, Lanfen Lin, Ruofeng Tong, Hongjie Hu, Qiaowei Zhang, Yutaro Iwamoto, Xianhua Han, Yen-Wei Chen, and Jian Wu.
\newblock Unet 3+: A full-scale connected unet for medical image segmentation, 2020.
\newblock URL \url{https://arxiv.org/abs/2004.08790}.

\bibitem[Rahman et~al.(2024)Rahman, Munir, and Marculescu]{rahman2024emcad}
Md~Mostafijur Rahman, Mustafa Munir, and Radu Marculescu.
\newblock Emcad: Efficient multi-scale convolutional attention decoding for medical image segmentation.
\newblock In \emph{Proceedings of the IEEE/CVF Conference on Computer Vision and Pattern Recognition}, pages 11769--11779, 2024.

\bibitem[Hung et~al.(2023)Hung, Manh, Oanh, Thuy, and Sang]{hung2023ugcanet}
Pham~Vu Hung, Nguyen~Duy Manh, Nguyen~Thi Oanh, Nguyen~Thi Thuy, and Dinh~Viet Sang.
\newblock Ugcanet: A unified global context-aware transformer-based network with feature alignment for endoscopic image analysis.
\newblock \emph{arXiv preprint arXiv:2307.06260}, 2023.

\bibitem[Bui et~al.(2024)Bui, Hoang, Nguyen, Tran, and Le]{bui2024meganet}
Nhat-Tan Bui, Dinh-Hieu Hoang, Quang-Thuc Nguyen, Minh-Triet Tran, and Ngan Le.
\newblock Meganet: Multi-scale edge-guided attention network for weak boundary polyp segmentation.
\newblock In \emph{Proceedings of the IEEE/CVF Winter Conference on Applications of Computer Vision}, pages 7985--7994, 2024.

\bibitem[Ahmed et~al.(2023)Ahmed, Fahim, Islam, Islam, and Shatabda]{ahmed2023dolg}
Md~Rayhan Ahmed, Md~Asif~Iqbal Fahim, AKM~Muzahidul Islam, Salekul Islam, and Swakkhar Shatabda.
\newblock Dolg-next: Convolutional neural network with deep orthogonal fusion of local and global features for biomedical image segmentation.
\newblock \emph{Neurocomputing}, 546:\penalty0 126362, 2023.

\end{thebibliography}


\end{document}